\documentclass[twocolumn,twocolappendix,times]{aastex63}

\usepackage{breqn}

\newcommand\angstrom{\AA}

\shorttitle{The Peculiar Blazar PKS\,1413+135}
\shortauthors{Readhead et al.}

\graphicspath{{./}{figures/}}

\begin{document}

\NewPageAfterKeywords 

\title{The  Relativistic Jet Orientation and Host Galaxy of the Peculiar Blazar PKS\,1413+135}

\correspondingauthor{Anthony Readhead}
\email{acr@caltech.edu}

\author{A.C.S Readhead}
\affiliation{Owens Valley Radio Observatory, California Institute of Technology,  Pasadena, CA 91125, USA}
\author{V. Ravi}
\affiliation{Owens Valley Radio Observatory, California Institute of Technology,  Pasadena, CA 91125, USA} 
\author{I. Liodakis}
\affiliation{Kavli Institute for Particle Astrophysics and Cosmology, Department of Physics,20
 Stanford University, Stanford, CA 94305, USA}
\author{M.L. Lister}
\affiliation{Department of Physics and Astronomy, Purdue University, 525 Northwestern Avenue, West Lafayette, IN 47907, USA}
\author{V. Singh}
\affiliation{Astronomy and Astrophysics Division, Physical Research Laboratory, Ahmedabad 380009, India}
\author{M.F. Aller}
\affiliation{2 Department of Astronomy, University of Michigan, 311 West Hall, 1085 S. University Avenue, Ann Arbor, MI 48109, USA}
\author{R. D. Blandford}
\affiliation{Kavli Institute for Particle Astrophysics and Cosmology, Department of Physics,20
 Stanford University, Stanford, CA 94305, USA}
\author{I. W. A. Browne}
\affiliation{Jodrell Bank Centre for Astrophysics, University of Manchester, Oxford Road, Manchester M13 9PL, UK} 

 \author{V. Gorjian}
\affiliation{Jet Propulsion Laboratory, California Institute of Technology, Pasadena, CA 91109, USA}
 \author{K.J.B. Grainge}
\affiliation{Jodrell Bank Centre for Astrophysics, University of Manchester, Oxford Road, Manchester M13 9PL, UK}
\author{M. A. Gurwell}
\affiliation{Center for Astrophysics $|$ Harvard \& Smithsonian, 60 Garden Street, Cambridge MA 02138 USA }
\author{M.W. Hodges}
\affiliation{Owens Valley Radio Observatory, California Institute of Technology,  Pasadena, CA 91125, USA}
 
 \author{T. Hovatta}
\affiliation{Finnish Centre for Astronomy with ESO (FINCA), University of Turku, FI-20014 University of Turku, Finland}
\affiliation{Aalto University Mets\"ahovi Radio Observatory,  Mets\"ahovintie 114, 02540 Kylm\"al\"a, Finland}
 \author{S. Kiehlmann}
\affiliation{Department of Physics and Institute of Theoretical and Computational Physics, University of Crete, 71003 Heraklion, Greece} \author{A. L\"ahteenm\"aki}
\affiliation{Aalto University Mets\"ahovi Radio Observatory,  Mets\"ahovintie 114, 02540 Kylm\"al\"a, Finland} 
\affiliation{Aalto University Department of Electronics and Nanoengineering, PO Box 15500, 00076 Aalto, Finland}
\author{T. Mcaloone}
\affiliation{Jodrell Bank Centre for Astrophysics, University of Manchester, Oxford Road, Manchester M13 9PL, UK} 
\author{W. Max-Moerbeck} 
\affiliation{Departamento de Astronomía, Universidad de Chile, Camino El Observatorio 1515, Las Condes, Santiago, Chile}

 \author{V. Pavlidou} 
\affiliation{Department of Physics and Institute of Theoretical and Computational Physics, University of Crete, 71003 Heraklion, Greece}

\author{T. J. Pearson}
\affiliation{Owens Valley Radio Observatory, California Institute of Technology,  Pasadena, CA 91125, USA}
\author{A. L. Peirson}
\affiliation{Kavli Institute for Particle Astrophysics and Cosmology, Department of Physics,20
 Stanford University, Stanford, CA 94305, USA}
\author{E. S. Perlman}
\affiliation{Dept. of Physics and Space Sciences, Florida Institute of Technology, 150 W. University Boulevard, Melbourne, FL, 32901, USA}
\author{R.A. Reeves}
\affiliation{CePIA, Astronomy Department, Universidad de Concepci\'on,  Casilla 160-C, Concepci\'on, Chile} 
\author{B. T. Soifer}
\affiliation{Spitzer Science Center, Mail Stop 314-6, California Institute of Technology, Pasadena, CA 91125, USA}
\author{G.B. Taylor}
\affiliation{Department of Physics and Astronomy, University of New Mexico, Albuquerque, NM 87131, USA}
\author{M. Tornikoski}
\affiliation{Aalto University Mets\"ahovi Radio Observatory,  Mets\"ahovintie 114, 02540 Kylm\"al\"a, Finland}
\author{H.K. Vedantham}
\affiliation{ASTRON, Netherlands Institute for Radio Astronomy, P.O. Box 2, 7990 AA Dwingeloo, the Netherlands}
\author{M. Werner}
\affiliation{Jet Propulsion Laboratory, California Institute of Technology, Pasadena, CA 91109, USA}
\author{P.N. Wilkinson}
\affiliation{Jodrell Bank Centre for Astrophysics, University of Manchester, Oxford Road, Manchester M13 9PL, UK}

\author{J. A. Zensus}
\affiliation{Max-Planck-Institut f\"ur Radioastronomie, Auf dem H\"ugel 69, D-53121 Bonn, Germany}

\begin{abstract}

 PKS\,1413+135 is one of the most peculiar blazars known. Its strange properties led to the hypothesis almost four decades ago that it is gravitationally lensed by a mass concentration associated with an intervening galaxy.  It  exhibits symmetric achromatic variability, a rare form of variability that has been attributed to gravitational milli-lensing. It has been classified as a BL Lac object, and is one of the rare objects in this class with a visible counterjet.  BL Lac objects have jet axes aligned close to the line of sight. It has also been classified as a compact symmetric object, which have jet axes not aligned close to the line of sight. Intensive efforts to understand this blazar have hitherto failed to resolve even the questions of the orientation of the relativistic jet, and the host galaxy. Answering these two questions is important as they challenge our understanding of jets in active galactic nuclei and the classification schemes we use to describe them.  We show that the jet axis is aligned close to  the line of sight and PKS\,1413+135 is almost certainly not located in the apparent host galaxy, but is a background object in the redshift range $0.247 <  z < 0.5$. The intervening spiral galaxy at $z = 0.247$ provides a natural host for the putative lens responsible for symmetric achromatic variability and is shown to be a Seyfert 2 galaxy. We also show that, as for the radio emission, a ``multizone'' model is needed to account for the high-energy emission.

\end{abstract}


\section{Introduction}\label{sec:intro}

Blazars are a class of relativistic jetted active galactic nuclei (jetted-AGN) in which the jet axis is aligned close to the line of sight \citep{1966Natur.211..468R,1980ARA&A..18..321A,2019ARA&A..57..467B}. The radio source PKS\,1413+135 is one of the most puzzling blazars known, largely  due to uncertainties about its jet orientation relative to the line of sight and about its host galaxy, which have persisted now for almost four decades \citep[see, e.g.,][]{2002AJ....124.2401P}. 
Blazars are classified as either flat-spectrum radio quasars (FSRQs) or BL~Lac objects (BL~Lacs), depending on whether their optical spectra are those of an AGN with strong emission lines, or virtually featureless.  PKS\,1413+135 has been classified as a BL~Lac object based on its optical spectrum, which was thought to be virtually featureless \citep{1981Natur.293..714B,1981Natur.293..711B} although a narrow weak [\ion{O}{2}] line with redshift $z=0.247$ was found by \citet{1992ApJ...400L..17S} associated with the apparent host, which is an edge-on spiral galaxy \citep{1991MNRAS.249..742M}.  

Jetted-AGN are rarely associated with spiral host galaxies. Yet, as shown by \citet{2002AJ....124.2401P},  PKS\,1413+135 is projected on the sky  $13\pm 4 $\,mas ($52\pm 16 $ pc)  from the isophotal center of this spiral.
PKS\,1413+135  must be either located in the spiral  or  a background source, but it has heretofore proven impossible to establish which of these possibilities applies \citep{1981Natur.293..714B,1989LNP...334...64S,1991MNRAS.249..742M,1992ApJ...400L..13C,1994MNRAS.268..681M, 1994ApJ...424L..69P,1996AJ....111.1839P,1996ApJS..103..109W,1999MNRAS.309.1085L,2002AJ....124.2401P,2017ApJ...845...89V}.

As a blazar and a BL~Lac object, PKS\,1413+135 is expected to be oriented with its jet axis almost along the line of sight \citep{1966Natur.211..468R,1978bllo.conf..328B,1978Natur.276..768R}. But based on its radio structure, PKS\,1413+135 has been classified as a compact symmetric object (CSO) \citep{1994ApJ...432L..87W,1996ApJ...460..612R,1994ApJ...424L..69P,1996AJ....111.1839P,2002AJ....124.2401P,2010ApJ...713.1393W} -- CSOs are a class of jetted-AGN in which the jet axes are aligned closer to the sky plane than to the line of sight. 

The  conventional interpretation of BL~Lac objects has been challenged by \citet{1985Natur.318..446O}, who suggested that gravitational lensing, and not relativistic beaming, may be responsible for the strong variability in some BL~Lac objects. A number of authors have suggested that  PKS\,1413+135 is a gravitationally lensed BL~Lac \citep{1992ApJ...400L..17S,1994ApJ...424L..69P,1996AJ....111.1839P,1996ApJS..103..109W,2002AJ....124.2401P,2017ApJ...845...89V}.  

Uncertainty has therefore arisen over  the orientation of the jet axis in PKS\,1413+135 relative to the line of sight, the location of PKS\,1413+135 relative to the spiral galaxy,  and the possible roles of gravitational lensing versus relativistic beaming in PKS\,1413+135.

\defcitealias{2017ApJ...845...89V}{Paper~1}

Astrophysics is a phenomenologically rich field and there are many curious beasts in the cosmic menagerie. Nevertheless it is somewhat surprising that in addition to the catalog of peculiarities exhibited by this strange blazar,  \citet{2017ApJ...845...89V} (hereafter \citetalias{2017ApJ...845...89V})  discovered symmetric achromatic variability (SAV) in this object. SAV is a new and very rare form of AGN variability, and PKS\,1413+135 was just one of $\sim$ 1800 objects being monitored in a large flux-density monitoring program (Richards et al. 2011).

\begin{figure}
    \centering
    \includegraphics[width=0.99\linewidth]{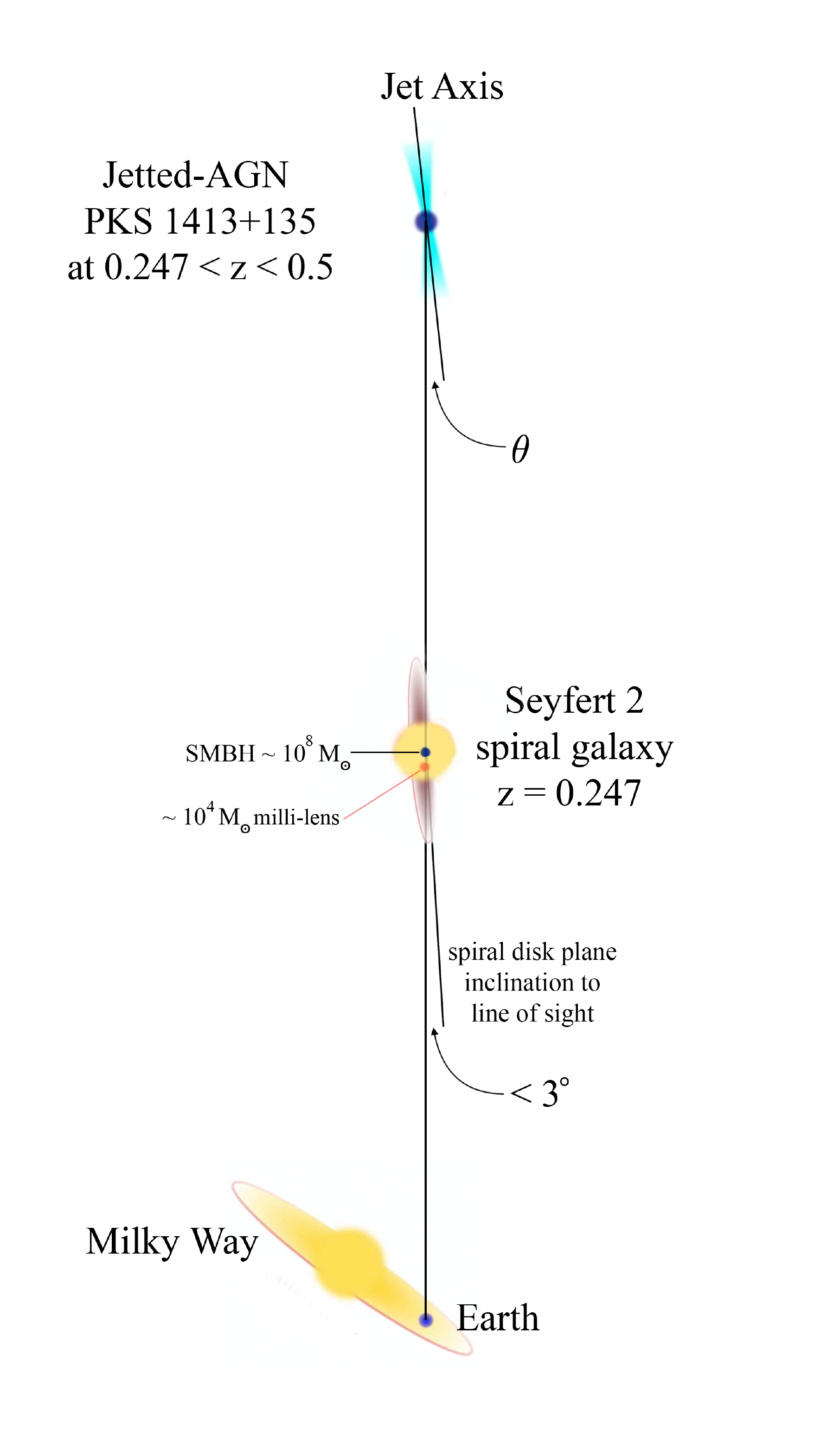}
    \caption{The orientations and configuration,  as determined in this paper, of the blazar PKS\,1413+135  and the Seyfert 2 spiral active galaxy at $z = 0.247$. The angle $\theta$ indicates the angle between the jet axis and the line of sight.  The blazar is highly variable at radio, infrared, X-ray, and $\gamma$-ray frequencies. Three gravitational lensing systems are considered: (i) the nuclear bulge of the spiral galaxy; (ii) the 
    SMBH powering the spiral AGN; and (iii) the putative milli-lens responsible for SAV (see text).}
    \label{cartoon}
\end{figure}

In this paper, we use multi-epoch very long baseline interferometry (VLBI), radio monitoring observations, and the infrared spectrum of PKS\,1413+135  to show that the jet axis is aligned close to the line of sight, and that the jetted-AGN is almost certainly a background source at $z < 0.5$ and not located in the spiral galaxy at $z = 0.247$, as shown in Fig. \ref{cartoon}. We also show that the variability in PKS\,1413+135 is a combination of intrinsic variability and SAV events, which may well be due to gravitational milli-lensing.

Since the comprehensive study of \citet{2002AJ....124.2401P} a significant amount of new observational data have been obtained on PKS\,1413+135, that we draw on in this paper:
\vskip 6pt
\noindent
1. It has been extensively studied in the MOJAVE VLBI monitoring program \citep{2009AJ....138.1874L,2019ApJ...874...43L}. MOJAVE monitoring shows that the components in the jet of of PKS\,1413+135 have unusually low apparent speeds for a bright blazar, with most of them being subluminal and  $v_{\rm app,max} = 1.72 \pm 0.11 c$, assuming that the radio source is at $z = 0.247$. If the radio source is at $z = 0.5$ then these values must be  increased by a factor of $1.58$ (see \S\ref{bulge}).
\vskip 6pt
\noindent
 2. It has been monitored with high cadence at centimeter, millimeter and sub-millimeter wavelengths.
\vskip 6pt
\noindent
3. It has been readily detected and 
monitored by \textit{Fermi}-LAT at $\gamma$-ray energies since 2008 \citep{2020ApJS..247...33A,2020ApJ...892..105A}.
\vskip 6pt
\noindent
4. An infrared \textit{Spitzer} spectrum has been obtained \citep{2010ApJ...713.1393W}.
\vskip 6pt
\noindent
5. An optical spectrum showing multiple narrow emission lines was presented in \citetalias{2017ApJ...845...89V}.

This paper is organized as follows: in \S\ref{active} we show the spiral galaxy to be an active galaxy and determine the mass of its SMBH; in \S\ref{morph} we discuss the radio structure of PKS\,1413+135 with particular emphasis on the jet and counterjet; in \S \ref{3scenarios} we discuss three scenarios of gravitational lensing in the combined system of PKS\,1413+135 and the spiral galaxy; in \S \ref{varradgamma} we discuss the variability of PKS\,1413+135 across the electromagnetic spectrum and show that the jet is closely aligned with the line of sight;  in \S \ref{coreslm}  we discuss superluminal motion in PKS\,1413+135 and refine our estimate of the jet alignment with the line of sight; 
in \S \ref{assoc} we discuss the location of PKS\,1413+135 and show that it is almost certainly not located in the spiral galaxy at z = 0.247 but is a background source; 
in \S \ref{other} we discuss other properties of PKS\,1413+135 that need to be explained;  and we conclude in \S \ref{discussion} with a brief summary of our findings.

Appendix~A presents some gravitational lensing formulae used in \S \ref{3scenarios}, and Appendix~B summarizes results, several of which have not been published before or have hitherto not been published for the case of a $\Lambda$CDM cosmology, on incoherent synchrotron radiation, relativistic beaming, and the variability Doppler factor used in \S \ref{varradgamma} and \S \ref{coreslm}.

In this paper we assume the following cosmological parameters: $H_0 = 71$\,km\,s$^{-1}$\,Mpc$^{-1}$, $\Omega_{\rm m} = 0.27$, $\Omega_\Lambda = 0.73$ \citep{2009ApJS..180..330K}.
 
\begin{figure*}
    \centering
    \includegraphics[width=0.65\linewidth]{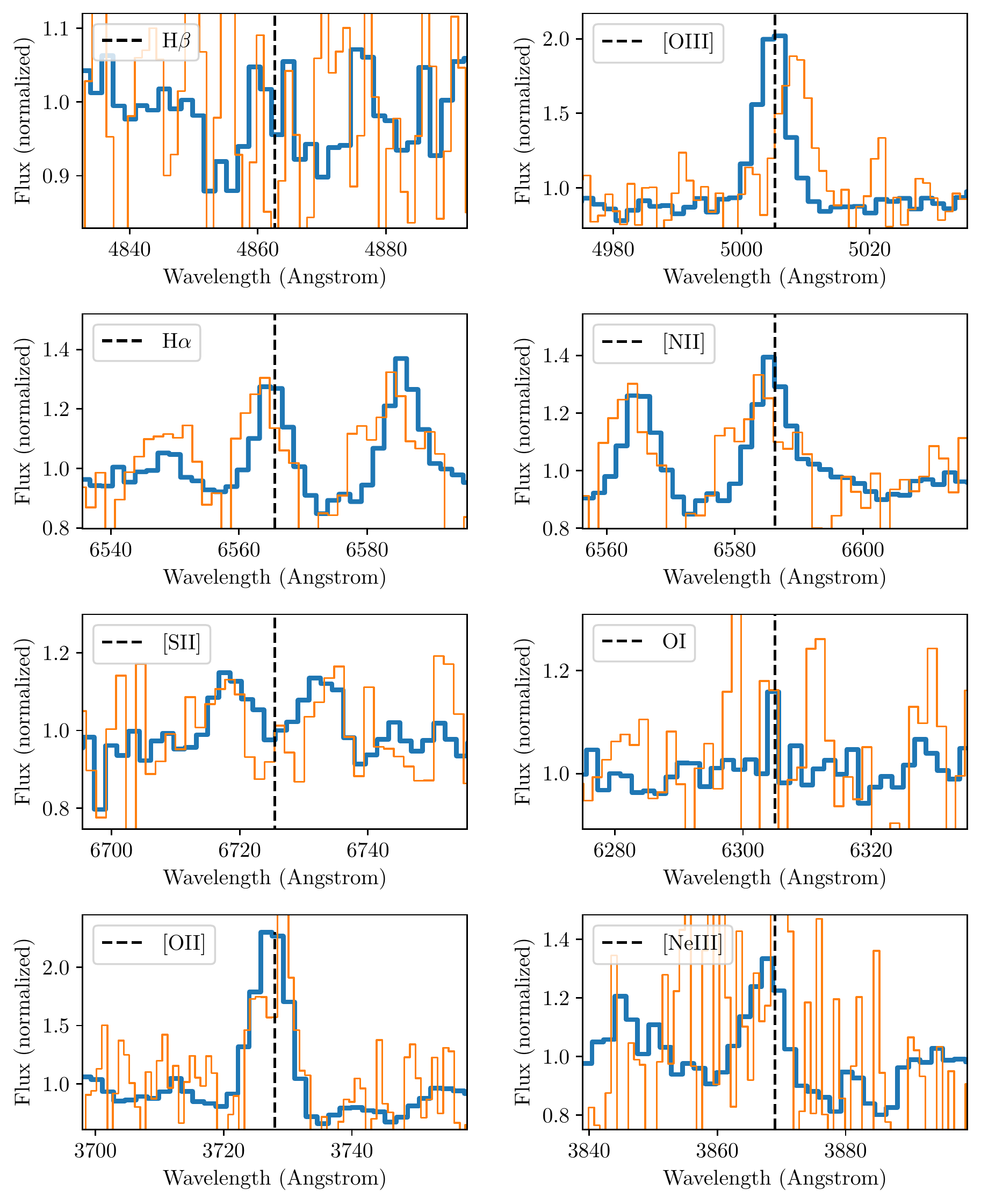}
    \caption{Spectral lines in the spiral galaxy at $z = 0.24675$ from the Keck\,I/LRIS spectrum (thick blue lines) and the SDSS/BOSS spectrum (thin orange lines). Both are continuum normalized, and plotted in rest wavelength.}
    \label{Keck}
\end{figure*}

\begin{figure*}
    \centering
    \includegraphics[width=0.7\linewidth]{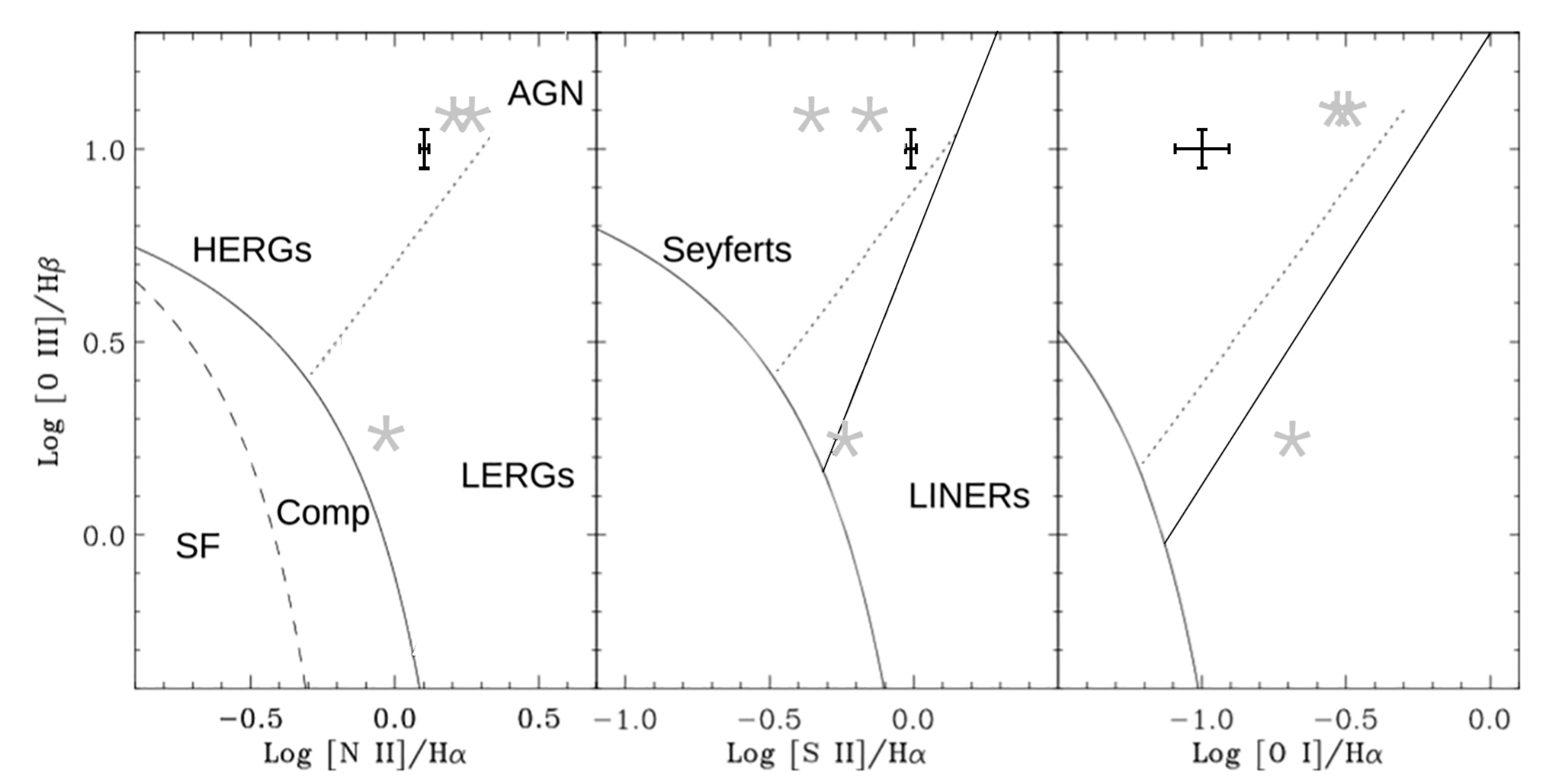}
    \caption{Emission line ratio diagnostics based on the corresponding figures from \citet{2006MNRAS.372..961K} and \citet{2015MNRAS.454.1556S}. The emission line ratios we have measured for PKS\,1413+135 are shown in black (error bars are $\pm1\sigma$). Also shown are the double-lobed radio sources associated with spiral galaxies discovered by \citet{2015MNRAS.454.1556S} (grey asterisks). The regions occupied by AGN and star forming and composite galaxies are separated by the solid curve, with the latter two being separated by the dashed curve. Seyfert galaxies lie to the left of the straight solid line, and LINERs lie to the right of this line. The dotted lines divide high-excitation and low-excitation radio galaxies.}
    \label{lineratios}
\end{figure*}

\section{Optical Properties of the Spiral Galaxy at z=0.247}\label{active}

In this section we discuss properties of the spiral galaxy that help to determine whether the radio source PKS\,1413+135 is associated with the galactic nucleus of the spiral or is a background object.

As revealed by the Hubble Space Telescope (HST) observations of \citet{1994MNRAS.268..681M}, the angle of inclination, $\xi$, between the plane of the spiral galaxy and the plane of the sky,
or equivalently the angle between the spin axis of the spiral galaxy
and the line of sight, is $\xi\ge 87\degr$.  So the angle
between the plane of the spiral disk and the line of sight is $\le 3\degr$,
i.e., this is truly an edge-on system, as indicated in Fig. \ref{cartoon}.

The host galaxies of blazars are almost exclusively giant elliptical galaxies   \citep{1995PASP..107..803U, 1997ApJ...479..642B, 1998ApJ...495..227L, 1999MNRAS.308..377M, 2003MNRAS.346.1055K}. However a small number of jetted-AGN are known to have spiral host galaxies, many of which are thought to have undergone interactions or recent merger events \citep{1980ApJ...242..894C,1983A&A...128..318N,1984MNRAS.209P..15U,1996ApJS..103..109W,1998MNRAS.299.1159A,1998ApJ...495..227L,2000ApJ...532..816U,2002MNRAS.336..319A,2010IAUS..267..119M,2011MNRAS.417L..36H,2014ApJ...788..174B,2015MNRAS.446.4176M,2015MNRAS.454.1556S}.

Prior to discussing the optical spectrum of the spiral galaxy, it is interesting to note the similarities between the spectra of Seyfert 2 galaxies and narrow-line radio galaxies in general \citep{1978ApJ...223...56K}, and that blazars generally exhibit broad emission lines, or exceedingly weak lines as in BL~Lac objects.

\subsection{The Optical Spectrum of the Spiral Galaxy}\label{optspectrum}

The Keck\,I low-resolution imaging spectrometer (LRIS)  \citep{1995PASP..107..375O} optical spectrum of the spiral %at $z=0.247$ 
was presented in Fig.~3 of \citetalias{2017ApJ...845...89V}. It exhibits several narrow emission lines at a redshift of $z=0.24675$. We analyzed the ratios of the line equivalent widths to determine the nature of the source of ionizing radiation in the galaxy. The spectrum was obtained with a 1\arcsec~slit at a slit position angle of $-70\degr$ (east of north), aligned with the mean parallactic angle of the observation. Much of the galaxy light was captured in the slit, as the galaxy major axis was only misaligned by $\sim30^{\circ}$. The effective spectral resolution of the observation (full-width at half-maximum, FWHM), as measured from night-sky emission lines, was a few hundred km\,s$^{-1}$ at  redshift $z=0.247$. An extinction correction of $A_{V}=0.067$ was applied to the spectrum using the \citet{1989ApJ...345..245C} extinction curve with $R_{V}=3.1$. This corresponds to the expected Milky Way foreground extinction \citep{2011ApJ...737..103S}. No attempt was made to correct for extinction within the spiral galaxy or the host of PKS\,1413+135. 

We analyzed the spectrum taken with the Keck\,I/LRIS presented in \citetalias{2017ApJ...845...89V}  as follows  (see Fig.~\ref{Keck}). Equivalent widths of the emission lines  measured by integrating the line profiles at the known wavelengths are shown in Table~\ref{eqwidths}. The apertures for the integrations were chosen to be 14\,{\angstrom}, which is twice the FWHM of the instrumental resolution. Smaller apertures did not include the full line widths, and larger apertures did not increase the measured equivalent widths. In the case of the [\ion{S}{2}] doublet a 30\,{\angstrom} aperture was used. The ${\rm H}\beta$ line was found to lie within an absorption trough, which was modeled as part of the continuum. The uncertainties in the equivalent widths were measured by bootstrap sampling from adjacent portions of the continuum. These uncertainties were propagated using standard techniques to the quoted uncertainties in the estimated line ratios discussed below. 

\begin{deluxetable}{l@{\hskip 8mm}ccc}
\tablecaption{Line equivalent widths in the spiral galaxy at $z = 0.247$\label{tab:xyz10}}
\tablehead{ Line& Wavelength&Equivalent width & Error\\
&{\angstrom}&{\angstrom}&{\angstrom}}
%\decimalcolnumbers
\startdata
$[\hbox{\ion{O}{2}}]$  & 3726, 3729 & 7.81 &0.08 \\ 
$[\hbox{\ion{Ne}{3}}]$ & 3868 & 2.13 &0.13 \\ 
${\rm H}\beta$  & 4861 & 0.473 &0.053 \\ 
$[\hbox{\ion{O}{3}}]$  & 5007 & 4.57 &0.05 \\ 
$[\hbox{\ion{O}{1}}]$  & 6300 & 0.158 &0.033 \\ 
${\rm H}\alpha$  & 6563 & 1.48 &0.03 \\ 
$[\hbox{\ion{N}{2}}]$  & 6583 & 1.88 &0.03 \\ 
$[\hbox{\ion{S}{2}}]$  & 6716, 6731 & 1.45 &0.04 \\ 
\enddata
\tablecomments{From the spectrum given in \citetalias{2017ApJ...845...89V}, taken with Keck\,I/LRIS on 2016 April 10.}
\label{eqwidths}
\end{deluxetable}

We detected several lines suitable for standard diagnostics of nuclear and star-formation activity \citep{1981PASP...93....5B,2006MNRAS.372..961K} (see Figs.~\ref{Keck} and \ref{lineratios}). In particular, we measured the following standard line ratios, all based on the positive line detections shown in Table~\ref{eqwidths}: 
$\log({\rm [ \hbox{\ion{O}{3}} ]/H}\beta) = 0.99 \pm 0.05$,
$\log({\rm [ \hbox{\ion{N}{2}} ]/H}\alpha) = 0.10 \pm 0.01$, 
$\log({\rm [ \hbox{\ion{S}{2}} ]/H}\alpha) = -0.01 \pm 0.02$, and 
$\log({\rm [ \hbox{\ion{O}{1}} ]/H}\alpha) = -1.0 \pm 0.1$. 
The ratios enable us to classify  the spiral at $z=0.247$ as a Seyfert galaxy. There is no hint of broad emission lines, and all lines are unresolved. The spectrum is therefore well matched to the Seyfert~2 class. We detect the [\ion{O}{2}] line along with the nearby [\ion{Ne}{3}] $\lambda3869$ line, which has a substantially higher ionization potential (41\,eV as compared to 13.6\,eV). The [\ion{O}{2}] line is an excellent tracer of ongoing star formation \citep{2004AJ....127.2002K}.  However, the line ratio
$\log({\rm [\hbox{\ion{Ne}{3}]/[\hbox{\ion{O}{2}}]}})$
is a reasonable diagnostic of AGN activity in metal-rich galaxies, but very few star-forming galaxies have ratios greater than our measured value of $-0.56\pm0.03$ regardless of metallicity \citep{2011ApJ...742...46T}. Finally, we measure a ratio of ${\rm H}\alpha/{\rm H}\beta=3.1\pm0.4$. This is consistent with the Balmer decrement for Case~B recombination under densities ($\sim10^{2}$--$10^{4}$\,cm$^{-3}$) and temperatures ($\sim10^{4}$\,K) typical of narrow-line AGN \citep{1989agna.book.....O}. We conclude that a radio-quiet AGN resides in the spiral galaxy. 

However, our observations raise a puzzle. Given the extremely large absorbing column towards the AGN in PKS\,1413+135 \citep{1992ApJ...400L..17S}, and the edge-on orientation of the galaxy, how is it that even the narrow-line region is not obscured? A clue is provided by the [\ion{O}{2}] line. Although this line is often associated with AGN activity, the low corresponding ionization potential means that it is only weakly excited in the narrow-line regions of AGN, and not at all in the broad-line regions \citep{2001AJ....122..549V}. Instead, it is often associated with more extended nebulosities bathed in the AGN continuum, known as extended emission-line regions (EELRs) \citep{1984ApJ...285L..49S,1986ASSL..121..297F,1987MNRAS.227...97R,2014MNRAS.443..755H}. These regions can extend from a few to over 100\,kpc from the nucleus, and are sometimes anisotropic, being associated with ionizing cones in AGN \citep{1991RPPh...54..579O}. 

The detection of several narrow emission lines in the optical spectrum of the spiral at $z=0.247$ is itself somewhat surprising, because previous observations by \citet{1992ApJ...400L..17S} with a CCD spectrograph detected only the [\ion{O}{2}] $\lambda3727$ emission-line doublet. Unfortunately the 1992 spectrum is not available and we cannot determine whether the spectrum may have changed. We inspected a publicly available Sloan Digital Sky Survey (SDSS) Baryon Oscillation Spectroscopic Survey (BOSS) spectrum of the spiral at $z=0.247$ obtained on 2012 March 25, and found that several of the lines in the Keck\,I/LRIS spectrum were also detected in the SDSS spectrum with the same equivalent widths as in our own observations (see Fig.~\ref{Keck}). The most likely explanation is that the higher excitation lines were simply too weak to detect with the sensitivity of the 1992 observations.

We find no evidence of extended emission in our two-dimensional spectrum of the spiral at $z=0.247$ at the position of the [\ion{O}{2}] line beyond the $\sim0.7\arcsec$  seeing smear, which corresponds to a 3\,kpc projected size. This is nonetheless consistent with the EELR hypothesis. Additionally, based on our detection of the [\ion{O}{1}] $\lambda 6300$ line, we measure a low value of 
$\log({\rm [\hbox{\ion{O}{1}}]/[\hbox{\ion{O}{3}}]}) = -1.5 \pm0.1$. 
This, together with the line ratios discussed above, is entirely consistent with the emission-line diagnostics of EELRs presented in \citet{1987MNRAS.227...97R}, but only marginally consistent with a nuclear narrow-line region.

\subsection{The Mass of the SMBH in the Spiral Galaxy}\label{SMBHmass}
 
We estimate the mass of supermassive black hole in the spiral galaxy using 

the empirical black hole mass -- bulge luminosity relations given by \citet{2013ApJ...764..184M} and \citet{2009ApJ...694L.166B}. \citet{2013ApJ...764..184M} present scaling relations for black hole mass and host galaxy properties for early-type and late-type galaxies. As a sanity check, we also use the empirical relation given in \citet{2009ApJ...694L.166B} for a sample of AGN. \citet{2009ApJ...694L.166B} use black hole mass measurements based on reverberation mapping and bulge luminosities from two-dimensional decompositions of HST images.

\begin{table}
\caption{Bulge and SMBH Mass Estimates\label{tab:smbh}}

\begin{tabular}{lll}
\hline
Parameter & Estimated value & Note\\
\hline
$m_{\rm V}^{\rm disk}$  &  18.9 \\
$m_{\rm V}^{\rm bulge}$ & 20.6 \\
$M_{\rm V}^{\rm bulge}$ & $-19.84$ \\
$L_{\rm V}^{\rm bulge}$ & $7.4 \times 10^{9} L_\odot$ \\
$\log_{10}(M_{\rm SMBH}/M_\odot)$  &  $7.97\pm0.25$  & $^{\rm a}$\\
$\log_{10}(M_{\rm SMBH}/M_\odot)$  &  $7.99\pm0.46$  & $^{\rm b}$\\
$\log_{10}(M_{\rm SMBH}/M_\odot)$  &  $7.88\pm0.07$  & $^{\rm c}$\\
\hline
\end{tabular}

\tablecomments{Estimates for the spiral galaxy at $z = 0.247$: The black hole mass ($M_{\rm SMBH}$) is estimated using black hole mass--bulge luminosity relations.  Errors shown on the mass estimates of the SMBH are based on the uncertainties in the scaling relations given below and do not take the intrinsic spread in the mass-luminosity relation into account (see text).}
\raggedright
\tablenotetext{a}{Using the \citet{2013ApJ...764..184M} relation for early-type galaxies: $\log(M_{\rm SMBH}/M_{\odot}) = \alpha + \beta \log(L_{\rm V}^{\rm bulge}/10^{11}L_{\odot}$)
where $\alpha = 9.23\pm0.10$, $\beta = 1.11\pm0.13$.}
\tablenotetext{b}{Using the \citet{2013ApJ...764..184M} the relation for late-type galaxies with $L_{\rm V}^{\rm bulge} \leq 10^{10.8} L_\odot$, the same form as (a) but with $\alpha = 9.10\pm0.23$, $\beta = 0.98\pm0.20$.}
\tablenotetext{c}{Using the \citet{2009ApJ...705..199B} relation for AGN: 
$\log(M_{\rm SMBH}/ M{_\odot}) = \alpha + \beta \log(L_{\rm V}^{\rm bulge}/10^{10}L_{\odot}$); 
$\alpha = 7.98\pm0.06$, $\beta = 0.80\pm0.09$.}

\label{mass}
\end{table}

\begin{figure*}
    \epsscale{1.17}
    \plotone{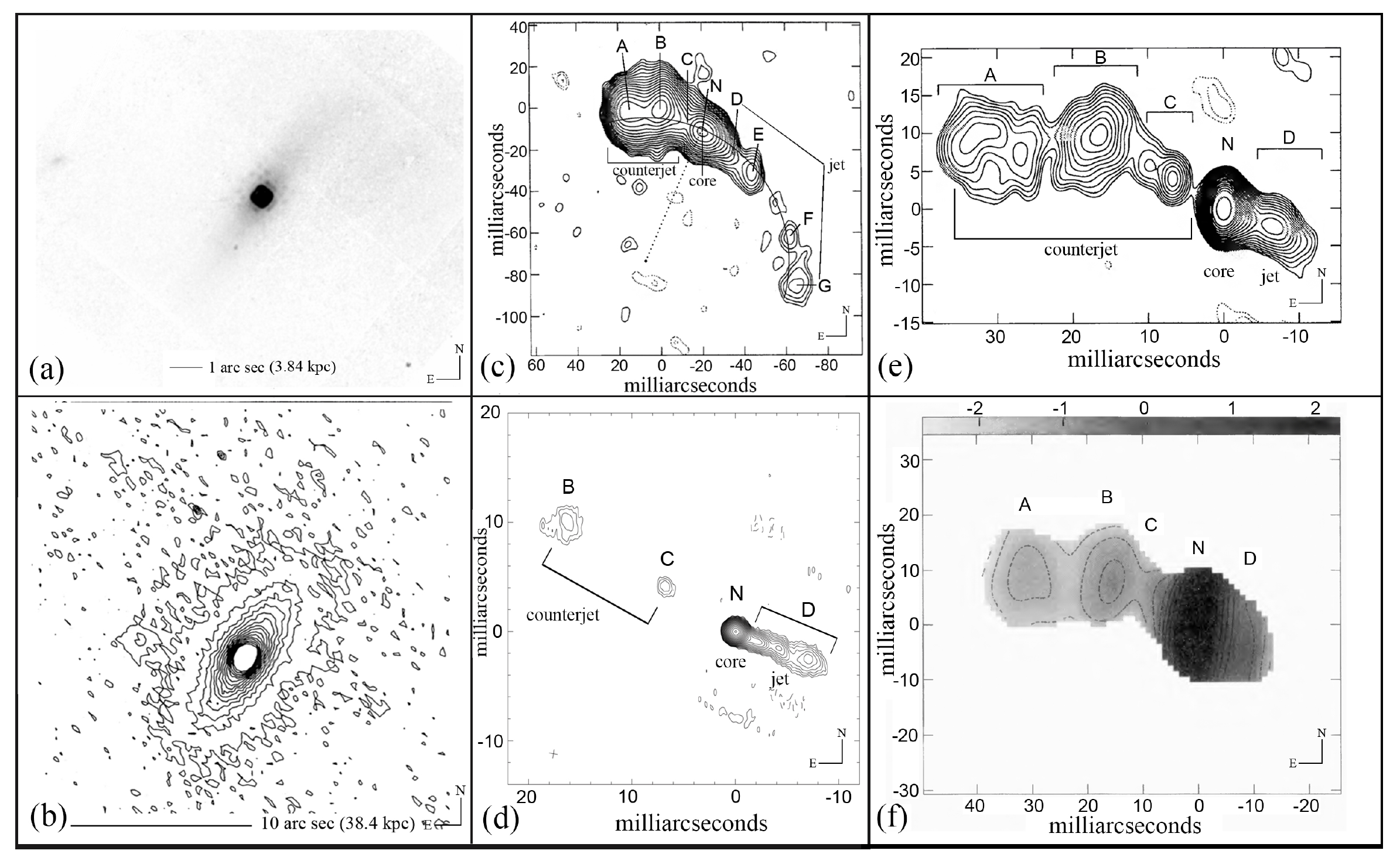}
    \caption{Six images of the PKS\,1413+135 and spiral galaxy system that show the near-orthogonality projected on the sky of the optical/IR spiral disk plane and the nuclear radio jet and counterjet: 
    (a)~Mosaicked HST NICMOS image of \citet{2002AJ....124.2401P}; (b)~$R$-band William Herschel Telescope image from \citet{1991MNRAS.249..742M}; 
    (c)~1.67\,GHz VLBI image from \citet{1996AJ....111.1839P} showing the total extent of the radio structure that has been observed. The contours are spaced by a factor of $\sqrt{2}$ in brightness. No features on scales greater than $\sim 110$\,mas have been observed. Component N marks the nucleus or radio core. The bright counterjet is designated by A, B and C. This image shows the curving structure of the radio emission. The inscribed arc is centered on a point $\sim 70$\,mas southeast of  the radio core N to illustrate the Einstein radius of a $\sim 10^9 M_\odot$ gravitational lens (see \S\ref{absence});  
    (d)~VLBI map made from stacking 15\,GHz MOJAVE images from 23 epochs between 1995 Jul 16 and 2011 May 26, showing the bright core and the emerging  jet~D and  inner counterjet component~C. The contours are spaced by a factor of $\sqrt{2}$ in brightness with the lowest positive contour at $0.243$\,mJy/beam (image from  \citealt{2017MNRAS.468.4992P} but with different contour levels).
    (e)~5\,GHz image from \citet{1996AJ....111.1839P} showing a clear bend in the counterjet at component~B. The contours are spaced by a factor of $\sqrt{2}$ in brightness.
    (f)~2.3\,GHz to 5\,GHz spectral index map from \citet{1996AJ....111.1839P}. 
    The letters designating components in panels (c--f) all refer to the same components.  }
    \label{map}
\end{figure*}

For the spiral at $z=0.247$, \citet{1991MNRAS.249..742M} showed that optical images in $R$ and $I$ bands clearly exhibit the presence of 
a small bulge while the galaxy disc dominates the emission in the $B$ and $V$ bands. 
Bulge$-$disk decomposition based on the surface brightness distributions gave the following magnitudes for the disk and bulge: $m_{\rm R}^{\rm disk} = 18.9$,  $m_{\rm R}^{\rm bulge} = 20.6$. At the redshift ($z = 0.2467$) of the spiral galaxy, $R$-band magnitudes in the observed frame approximately correspond to $V$-band magnitudes in the rest frame. We converted the bulge magnitude into a luminosity, which we then used  to estimate the mass of the SMBH. Table~\ref{mass} lists the  magnitude, the luminosity, and the SMBH mass estimates for the spiral at $z=0.247$.  

The uncertainties in the derived masses shown in Table~\ref{mass} are based on the uncertainties in the mean black hole mass -- bulge luminosity relations derived by \citet{2013ApJ...764..184M} and do not include the  intrinsic scatter in this relationship. From Fig.~2 of \citet{2013ApJ...764..184M}   the intrinsic scatter in the SMBH mass is
two orders of magnitude at a given bulge luminosity, and two-thirds of the points lie
within a band of width one order of magnitude.  Thus the 1-sigma uncertainty in ${\rm log}_{10} (M_{\rm SMBH}/M_\odot)$ is 0.5, and this is the best estimate of the uncertainty to use since it relies on no fitting and takes systematic errors into account automatically. Thus our estimate for the mass of the SMBH in this early type spiral galaxy is ${\rm log}_{10} (M_{\rm SMBH}/M_\odot) = 7.97\pm 0.5$. We note that a 2$\sigma$ variation on the low side would place the mass of the SMBH at $\sim 10^7 M_\odot$, and a 2$\sigma$ variation on the high side would place it at $\sim 10^9 M_\odot$. Therefore in our discussion of possible gravitational lensing by the SMBH in the spiral galaxy in \S \ref{absence} we consider the range of masses $10^7$--$ 10^9 M_\odot$ for the SMBH.

\section{The radio Structure of PKS\,1413+135}\label{morph}

\begin{figure*}
   \centering
    \includegraphics[width=0.98\linewidth]{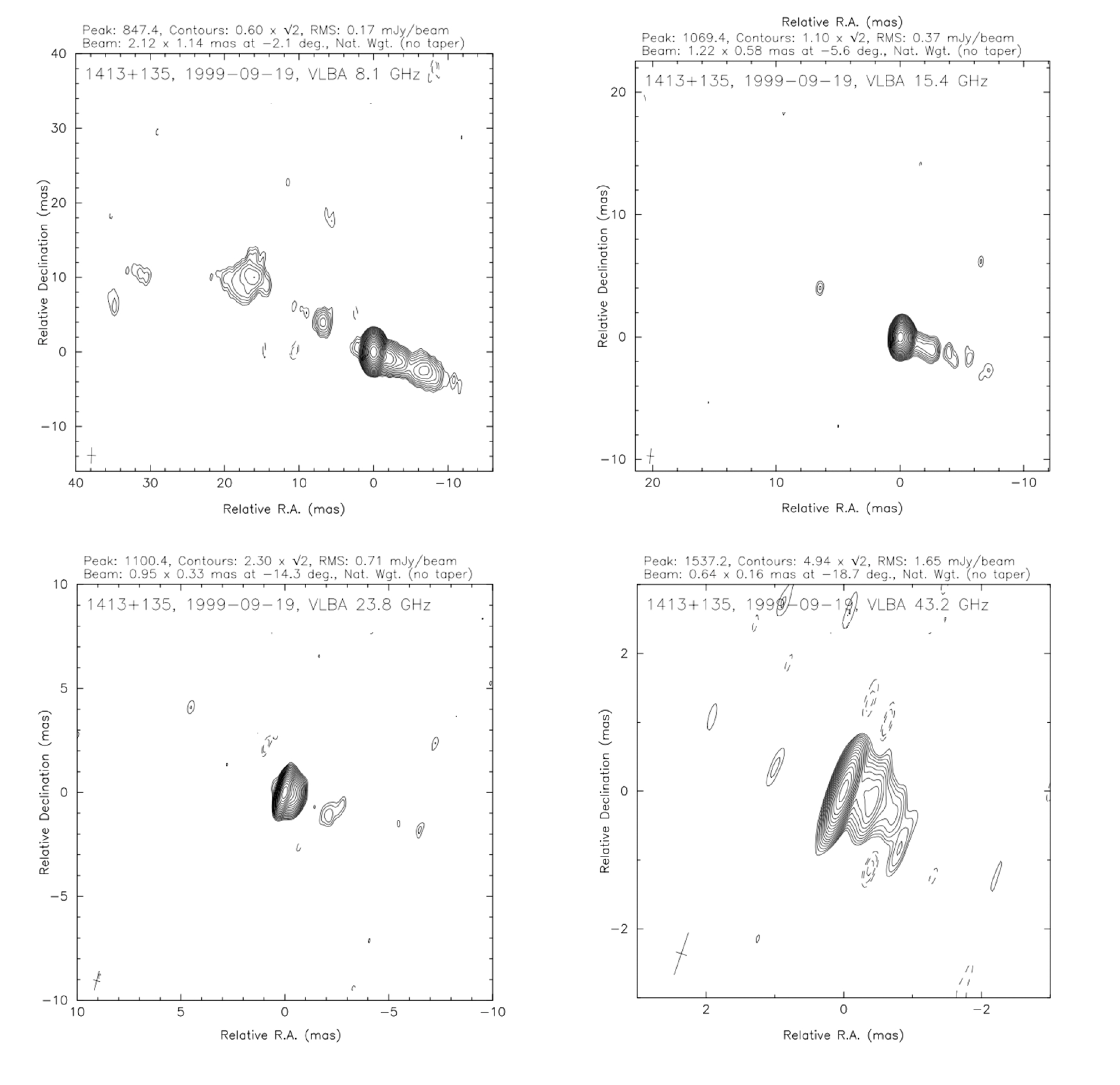}
    \caption{VLBI maps of PKS\,1413+135 at 8.1\,GHz, 15.4\,GHz, 23.8\,GHz, and 43.2\,GHz from observations made on 19 
    September 1999. The synthesized beam is indicated in the bottom left hand corner of each map. These maps show no evidence of gravitational lensing on scales from 1--40\,mas, and they also show no evidence of interstellar scattering (see text).}
    \label{VLBImaps}
\end{figure*}

In Fig.~\ref{map} we show near-infrared, optical, and radio images of the system (jetted-AGN PKS\,1413+135 and spiral galaxy). The low-frequency radio structure of PKS\,1413+135  at 1.67\,GHz is shown in Fig.~\ref{map}(c). This image shows the full extent of the radio structure that has been observed in this source. Four VLBI images from observations on 1999 September 19 at frequencies of 8.1, 15.4, 23.8, and 43.2\,GHz are shown in Fig.~\ref{VLBImaps}.

The near-orthogonality of the radio jet and the spiral disk plane in PKS\,1413+135 can be seen in Fig.~\ref{map}. The disk has position angle ${\rm PA} = 143\degr \pm 3\degr$.  The position angle of the innermost seven features in the radio jet of PKS\,1413+135 in the MOJAVE maps (Fig.~\ref{map}d)  is ${\rm PA} = 245.1\degr \pm 4\degr$. Thus the difference is $\Delta{\rm PA} = 102.1\degr \pm 5\degr$. So in the plane of the sky the innermost regions of the radio jets are aligned to within $12.1\degr \pm 5\degr$ with the normal to the disk projected onto the sky of the spiral galaxy, based on the near-infrared observations of \citet{2002AJ....124.2401P} and the MOJAVE VLBI observations of \citet{2016AJ....152...12L,2019ApJ...874...43L}.

Fig.~\ref{map}(c) shows that the radio source has overall size  $\sim 110$\,mas.  We show in \S \ref{bulge}, on the basis of the lack of multiple images on the arc-second scale, that the redshift of the radio source must be less than 0.5. This corresponds to an angular diameter distance of 1.255\,Gpc.  At this distance 110\,mas corresponds to 660\,pc.   We see, therefore, 
 that at radio frequencies PKS\,1413+135  is both symmetric in appearance and smaller than 1 kpc. This is why PKS\,1413+135 has been classified as a CSO.

\begin{figure}
    \centering
    \includegraphics[width=0.99\linewidth]{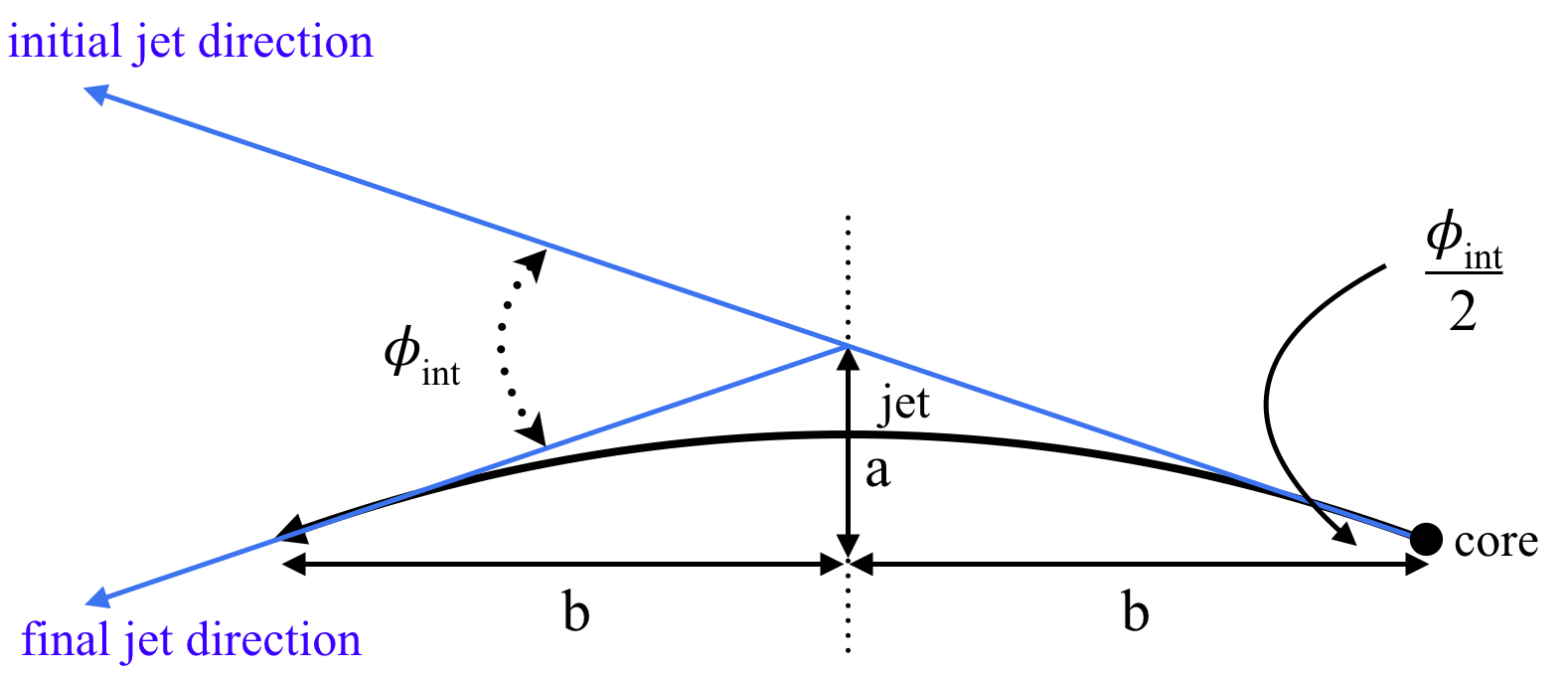}
    \caption{The geometry of the jet for the model we are assuming for PKS\,1413+135. The line of sight is perpendicular to the figure for $\theta=90\degr$. In discussing the curvature of the jet and counterjet we assume that the curvature is continuous across the core and continues on the counterjet side (see text). }
    \label{geometry}
\end{figure}

 However, the original intent of \citet{1994ApJ...432L..87W} in defining the CSO class was to discriminate between jetted-AGN with axes aligned closely with the line of sight and jetted-AGN with axes aligned closer to the sky plane than to the line of sight, which is not the case in PKS\,1413+135 (see Fig.~\ref{cartoon}). We show in this paper (\S \ref{varradgamma}) that the jet in PKS\,1413+135 is aligned close to the line of sight, and as such it should never have been classified as a CSO.  For this reason we do not consider this classification of PKS 1413+135 further in this paper. So the case of PKS\,1413+135 demonstrates that the existing CSO classification scheme has failed  in its original objective. For this reason the classification scheme for CSOs needs to be tightened to include variability, radio spectrum,  and apparent speed criteria.  This is discussed in detail in a separate paper (Kiehlmann et al., in preparation).

\subsection{The Jet and  Counterjet of PKS\,1413+135}\label{jetcjet}

As can be seen in the 8.1\,GHz image of Fig.~\ref{VLBImaps} and in the  MOJAVE stacked-epoch 15\,GHz image of Fig.~\ref{map}(d), the western jet has a fairly constant position angle of $249\degr$ for the first 10\,mas. At lower frequencies (see the 1.4\,GHz map of \citealt{1996AJ....111.1839P}), the jet arcs gradually towards PA = $180\degr$, where it fades out at 90\,mas from the core. There is thus direct evidence that the jet ridgeline in PKS\,1413+135 is intrinsically bent. This is further supported by the fact that the features in the eastern jet (out to 20\,mas in the stacked MOJAVE image, with PA = $59\degr$) are not co-linear with the western jet, but are offset by $10\degr$.  In contrast, two well-studied two-sided jetted AGN in the MOJAVE program, NGC\,1052 and Cygnus\,A, that lie close to the plane of the sky have a high degree of co-linearity in the PAs and proper motions of their approaching and receding jets \citep{2019ApJ...874...43L}.

To understand the apparent curvature of the jet in PKS\,1413+135, let us first assume that the jet has a small intrinsic bend, $\phi_{\rm int}$, of constant curvature in a single plane as shown in Fig.~\ref{geometry}. It is shown here for a bend in the sky plane, so here the angle between the jet axis and the line of sight is $\theta = 90\degr$. The real bend is only a few degrees, but has been exaggerated here for clarity.  If we rotate the jet about its mid-point, shown by the dotted line in Fig.~\ref{geometry}, to angle $\theta$ relative to the line of sight, length $a$ is unchanged and the projected length of side $b$ is $b \sin \theta$.  The observed bend in the jet is given by $\phi_{\rm obs}$ in the relation
\begin{equation}\label{geom}
\tan(\phi_{\rm obs} /2) = \tan(\phi_{\rm int} /2) / \sin\theta  \,. 
\end{equation}

In the case of PKS\,1413+135, as can be seen from Fig.~\ref{map}(c), the total observed bend along the jet from component N to component G is $\phi_{\rm obs} \sim 70\degr$, so equation~(\ref{geom}) yields $\tan(\phi_{\rm int} /2) \approx 0.7 \times \sin\theta$.
We show in \S \ref{varsav} that the angle between the jet axis and the line of sight is certainly $\le  7\degr$ and is very likely a few degrees. Thus in total we see that the observed bend  is $\sim 70\degr$ at angle 
$\theta \approx \phi_{\rm int} /1.4$. In other words an intrinsic bend in the jet of only a few degrees will produce the overall projected bend in the jet axis of $\sim 70\degr$ that we see in Fig.~\ref{map}(c) when viewed from within a few degrees of the jet axis. Of course the observed bend angle tends to $180\degr$ as $\theta \rightarrow 0$.

The simplest possible bending model is a monotonic bend that lies all in a single plane, as assumed above. Given a small viewing angle near the base of the jet, this would normally result in an apparent one-sided jet, due to the strong Doppler boosting of the approaching jet emission and de-boosting of the receding jet emission. The approaching jet would also rapidly fade with distance from the core as the viewing angle increases. However, if the bending plane lies perpendicular to the sky, the jet can cross our line of sight, resulting in apparent emission from the approaching jet on both sides of the core. Statistically, the odds of this alignment appear at first to be low, but in a flux-limited AGN sample, there is a strong preference for those jets that have maximum Doppler boosting factor. The latter occurs when a portion of the jet is pointing directly at us, which is the case for a jet crossing our line of sight. \citet[p99]{1999PhDT.........9L}  showed that in fact a simple parabolic bent jet will have a total apparent flux density much larger than a straight jet with identical inner jet viewing angle, provided the bending occurs in a plane nearly perpendicular to our line of sight. One notable case of this azimuthal favoritism is the blazar PKS\,1510$-$089, where \citet{2002ApJ...580..742H} provide strong evidence of a highly superluminal jet that makes a $180\degr$ apparent bend in the sky plane, with resulting co-linear jet emission on either side of the AGN core. We do, therefore, have to consider the possibility that all of the radio emission seen in this blazar is associated with the jet, but just happens to appear to straddle the nuclear core due to projection effects.

 One clear difference between the radio structures of PKS\,1413+135 and PKS\,1510$-$089 is the inner counterjet component~C in PKS\,1413+135 seen in Figs. \ref{map}(c,d,e,f). An even closer component of the inner counterjet, which is connected to the core component, is seen in the 8.1\,GHz and 23.8\,GHz images of Fig.\ref{VLBImaps}. No such inner counterjet, close to the core, is seen in PKS\,1510$-$089, and it is very hard to imagine that these two close-in components to the core, which are so well-aligned with the jet position angle projected across the core,  are actually components of the approaching curving jet that have been projected across the core to the counterjet side after the approaching jet curves through an apparent bend of $\sim 70\degr$ from components N through G as seen in Fig.~\ref{map} (c). There is also the surface brightness of the components to consider. The components within 10\,mas on the northeast side of the core have much higher surface brightness than components E, F, and G so it is extremely unlikely that they are further out along the jet than components E, F, and G.  In addition, as \citet{2002AJ....124.2401P} pointed out, the fact that the apparent counterjet undergoes a brightening at component B of Fig.~\ref{map}(e) just where the jet direction also changes, suggests that this is due to interaction with the interstellar medium.

For these reasons we interpret the bend in the counterjet 20\,mas northeast of the core at component~B as due to interaction with the surrounding medium and entrainment of material that is decelerating the outer sheath of the jet, giving rise to unbeamed emission from the sheath of the jet. Indeed, this interaction must begin much closer to the core to account for both component~C and the component of the counterjet that is attached to the core seen in the 8.1\,GHz and 23.8\,GHz images of Fig.~\ref{VLBImaps}, but the interaction must become stronger at the position of component~B.

 In PKS\,1413+135 the observed jet-to-counterjet flux-density ratio, $R$, is $<10$, so the counterjet is far brighter relative to the jet than usual for a blazar.   We interpret the unusual relative brightness of the counterjet, including components C, B, and A, as due to interaction with the ambient medium and deceleration of the jet.  Thus in our view the relatively low jet-to-counterjet flux-density ratio found in PKS\,1413+135 cannot be used to place constraints on the jet model. Furthermore this implies that the 60\% of the flux density at 1.67 GHz that comes from components A and B is isotropic.

Our main conclusions from this section are that the components southwest of the core are all jet components moving towards us, and the components northeast of the core are all counterjet components moving away from us, and for the remainder of this paper we adopt this interpretation.

\clearpage

\section{Three Gravitational Lensing Scenarios}\label{3scenarios}
 
If the radio source PKS\,1413+135 is not located in the spiral galaxy, its
 redshift is needed to interpret its observed properties.%
 so this is the next issue that we address.  
 We  determine the range of possible redshifts for the radio source based on gravitational lensing arguments, and we then also consider two other lensing scenarios that are of interest in this system. In  all three scenarios we assume that the radio source lies behind the spiral galaxy. We justify this assumption in \S \ref{assoc}.  The three scenarios we consider are (see Fig.~\ref{cartoon}): (i) gravitational lensing of the radio source by the nuclear bulge mass of the spiral galaxy; (ii) gravitational lensing of the background radio source by the $\sim 10^8M_\odot$ SMBH in the nucleus of the spiral galaxy; and (iii) gravitational milli-lensing of the background radio source by a $\sim 10^4 M_\odot$ mass condensate associated with the spiral galaxy.

The Einstein radius, $\Theta_{\rm E}$, gives the angular scale on which we would expect to see multiple images of the radio core due to gravitational lensing  \citep{1992ARA&A..30..311B}:
\begin{eqnarray}\label{einstein}
\Theta_{\rm E} &=& (4GM/c^2L)^{1/2} \nonumber \\
               &\approx& 90(M/10^3M_\odot)^{1/2}(L/1\, {\rm Gpc})^{-1/2}\;\mu{\rm as} .
\end{eqnarray}
Here $G$ is the gravitational constant and $L=L_{\rm l}L_{\rm s}/L_{\rm ls}$, where $L_{\rm l}$ is the distance to the lens, $L_{\rm s}$ is the distance to the source, and $L_{\rm ls}$ is the lens--source distance, all distances being angular diameter distances.\footnote{Note that we use upper case $\Theta$ to designate the angles in gravitational lensing and lower case $\theta$ to designate the angle between the jet axis and the line of sight.}
 
\subsection{Lensing by  the Nuclear Bulge Mass in the Foreground Spiral Galaxy}\label{bulge}
 
A piece of evidence that would appear to argue against PKS\,1413+135 being a background radio source is the absence of multiple images of the radio source  on arcsecond scales. Multiple images are expected  due to gravitational lensing by the mass associated with the nuclear bulge of the spiral galaxy.
\citet{1999MNRAS.309.1085L} have considered this question in detail and applied the method developed by \citet{1990MNRAS.243..192N} to derive a lower limit to the core radius of a lensing galaxy when no multiple images are seen. 
 They show that under the very conservative assumptions of a core radius of 2\,kpc and velocity dispersion of 180\,km\,s$^{-1}$ the redshift of the radio galaxy PKS\,1413+135 cannot be greater than $z = 0.5$. We  can therefore be confident that the redshift of the jetted-AGN PKS\,1413+135 lies in the range $0.247<z<0.5$ as indicated in Fig.~\ref{cartoon}.

\subsection{Lensing by  the SMBH in the Foreground Spiral Galaxy}\label{absence}
 
Given the above range in possible redshifts for the radio source, we calculate the Einstein radius for the  case $z = 0.5$, corresponding to the upper limit, and also for the case $z=0.25$, at which the radio source would lie at $\sim 1\%$ of the distance between $z = 0.247$ and $z = 0.5$. For the $z=0.5$ case, with the SMBH of $\sim 10^8M_\odot$ (see \S \ref{SMBHmass}) in the AGN of the spiral at redshift $z = 0.247$ acting as a gravitational lens, we have $L_{\rm l} = 793$\,Mpc,  $L_{\rm s} = 1.255$\,Gpc. Hence applying equation~(\ref{einstein}) we find  that $\Theta_{\rm E}=19.4$\,mas.  For the  $z = 0.25$ case, $L_{\rm s} = 800$\,Mpc, and the Einstein radius is $\Theta_{\rm E}= 3.0$\,mas.
  
  Since, as discussed in \S \ref{SMBHmass}, the scatter in the mass-luminosity relation is two orders of magnitude for early-type galaxies, we also consider the case where the mass of the SMBH is $\sim 10^7 M_\odot$.  In this case the Einstein radius with the radio source at $z=0.5$ is $\Theta_{\rm E}=6.1$\,mas. If the radio source is at $z = 0.25$,  we have  $\Theta_{\rm E}= 0.95$\,mas.
  
 The magnitude of the displacement between the radio core, which dominates the 15\,GHz emission from  the jetted-AGN PKS\,1413+135, and the SMBH at the centre of the spiral is critical to the question of whether or not we would expect to see multiple gravitational lens images of the radio core.   \citet{2002AJ....124.2401P} showed, on the basis of high-resolution HST near-infrared observations, that the near-infrared core of the blazar PKS\,1413+135 is offset by only $13\pm 4$\,mas from the centroid of the near-infrared isophotes  of the spiral galaxy, and in a direction perpendicular to the spiral disk and the dust lane, i.e., roughly in the direction of the jet. With the presence of a dust lane whose normal is slightly inclined to the plane of the sky there is the possibility of shifting the near-infrared spiral isophotes  either in the jet direction or in the opposite direction depending on whether the near side of the dust disk lies to the north-east or the south-west. Indeed, \citet{2002AJ....124.2401P} use this argument to suggest that the near-infrared spiral galaxy isophotes have been distorted by 13\,mas and that the real offset is zero, and therefore that the blazar is located in the spiral galaxy. The uncertainty in the offset obtained by  \citet{2002AJ....124.2401P} was $\pm 4$ milliarcseconds, so this would amount to a $\sim 3-\sigma$ systematic shift in the near-infrared isophotes.
 
 We are unaware of any direct evidence as to the orientation of the dust disk axis so it is quite possible that the shift is in the opposite direction by 13\,mas or more. The evidence that does exist comes from the VLBA \ion{H}{1} absorption measurements discussed by  \citet{2002AJ....124.2401P}. They find that the absorption seems to occur in front of the north-eastern radio emission. This suggests that the \ion{H}{1}, and possibly any associated dust, is in the foreground to the north-east of the spiral galaxy. This is in the right place to lead to an underestimate of the magnitude of the separation between the radio core and the spiral nucleus  rather than an overestimate and thus we think it is justified to assume in the following that the displacement might be as much as 26\,mas.
 
We therefore consider a range of projected offsets of the blazar from the SMBH of the spiral galaxy from 0 to 26\,mas, and a range of Einstein radii from 1 to 20\,mas. We will designate the projected angle between the radio core of PKS\,1413+135 and the SMBH in the spiral galaxy by the impact parameter $p$, and by $u$ when normalized by the Einstein radius, i.e. $u= p/\Theta_{\rm  E}$.  We will therefore consider the range $u=0 \rightarrow 26$.

 \begin{deluxetable*}{l@{\hskip 8mm}cccccc}
\tablecaption{Magnification Ratios and Separations \label{tab:xyz2}}
\tablehead{Magnification& Impact&Image+&Image$-$&Image+ and Image$-$ &Image+&FOV\\
$\;\;\;\;\;\;$Ratio&Parameter&Flux Density&Flux Density&Separation&Deflection&\\
&(Einstein radii)&(mJy)&(mJy)&(Einstein radii)&(Einstein radii)&(mas)}

\startdata
$\;\;\;\;\;\;\;\;\;$10  &1.215&915&92& 2.34 & $0.563$&94 \\ 
$\;\;\;\;\;\;\;\;\;$100  &2.846&915&9.2& 3.485 & 0.316&139 \\ 
$\;\;\;\;\;\;\;$ 1000  &5.4456&915&0.9& 5.80 & 0.178&232 \\
$\;\;\;\;\;\;\; 4.6\times 10^5 $ &26&915&0.002& 26.077 & 0.038&1043 \\
\enddata
\tablecomments{The two images produced of a point source by a point mass are designated by ``Image+'' and ``Image$-$''.  The magnification ratios indicated in the first column occur at the impact parameters shown in the second column. The corresponding separations of the two images, and deflection of Image+ from the true source position  are also shown.  The impact parameters, image separations and image deflection are all in units of the Einstein radius, $\Theta_{\rm E}$.  Also shown is the size of the field of view (in mas) that is required to see both images out to the corresponding impact parameter for  $\Theta_{\rm E} = 26$\,mas.}
\label{lensvals}
\end{deluxetable*}

Before we begin this discussion in detail, it is worth pointing out that, given that the range of the offset parameters $u=0 \rightarrow 26$  we  do not necessarily expect to see multiple images or other signatures of lensing due to the SMBH in the spiral galaxy in the radio images of the radio source PKS\,1413+135. So the absence of multiple radio images of the core does not rule out the background radio source hypothesis.
 
\begin{figure*}
    \centering
    \includegraphics[width=0.95\linewidth]{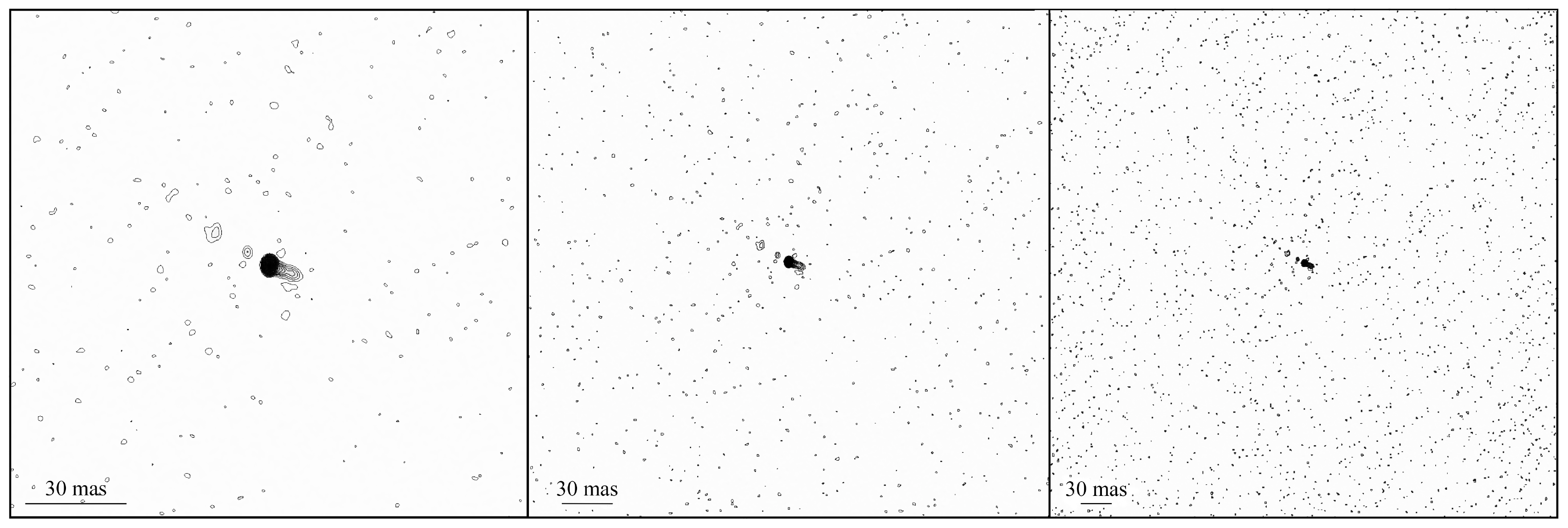}
    \caption{Widefield 15\,GHz VLBA image of PKS\,1413+135 from the MOJAVE 2001 epoch observations. Left panel: FOV 150 mas; Center panel: FOV 300 mas; Right panel: FOV 500 mas. Peak flux density 1.24\,Jy/beam, with the lowest contour 0.4\,mJy/beam and the second contour 1.2\,mJy/beam. Over the whole 500 mas field no second image of the core is visible  down to the level of 1.2\,mJy/beam, which corresponds to a magnification ratio of 1000 (see text).}
    \label{widefield}
\end{figure*}
 
 We now examine the maps in more detail.
 At first  sight the curving structure of the source seen in Fig.~\ref{map}(c), which has a radius of curvature of about 70\,mas, might appear to be explicable due to gravitational lensing by an intervening SMBH of mass $\sim 10^9 M_\odot$ located in the spiral galaxy. 
\citet{1996AJ....111.1839P} considered this possibility. They showed that the counterjet features (A, B, and C) have steep negative spectra, whereas the jet features (D and E) have inverted and flat spectra, respectively (see Fig.~\ref{map}f). \citet{1996AJ....111.1839P} were not able to rule out gravitational lensing by a $\sim 10^9 M_\odot$ SMBH because of the possibility that the steeper spectra of components A and B  were due to free-free absorption. However, as we have just seen, \citet{2002AJ....124.2401P} showed that the nucleus of the blazar is offset by at most 26\,mas  from the nucleus of the spiral galaxy, which is much too close to the core N (see Fig.~\ref{map}c) to account for the curvature. So gravitational lensing can be ruled out as an explanation of the overall curving structure. 
 
 We next consider the possibility of lensing on  smaller angular scales. The radio source PKS\,1413+135 is strongly dominated by the core at frequencies above 5\,GHz and the core is very bright down to 1.67\,GHz \citep{1996AJ....111.1839P}.
 
 For illustrative purposes, ignoring any convergence or shear due to the host galaxy, we consider gravitational lensing of a point source by a point mass. This produces two images.  The corresponding relationships between the separation of the two images, their magnification and the deflections of the images are given in Appendix~\ref{pointmass}. Using equations (\ref{Theta}) and (\ref{mu})  we have derived the parameters of interest, which are given in Table~\ref{lensvals}.  Here we show the normalized impact parameters, separations of the two images, and deflection of the brighter image from the true source position for magnification ratios of 10, 100 and 1000, as well as for the normalized maximum impact parameter of 26.  
 
 The core flux density of PKS\,1413+135 in the stacked image of Fig.~\ref{map}(d)
is 915\,mJy, and the lowest contour level is 0.243 mJy/beam, so the ratio of the core to the lowest contour is $\sim 3800$:1.  Given the noise visible on this map at the one-contour level we need a minimum of two to three contours to trust a detection, so the dynamic range of the map is $\sim 1000$.  
 
 On the 1\,mas scale, as we see from the 43.2\,GHz image in Fig.~\ref{VLBImaps}, there is no hint of multiple images of the flat-spectrum core, only steeper-spectrum components lying linearly along the jet and counterjet. In fact over the whole range from 1 to 40\,mas the morphology of PKS\,1413+135 seen in Fig.~\ref{map}(d) and Fig.~\ref{VLBImaps} simply does not look like multiple images of a gravitationally lensed object since the jet and counterjet are only gently curved within 20\,mas of the core. 
   
 The most stringent limits on possible secondary images of the radio core of PKS\,1413+135 are provided by the MOJAVE 15\,GHz observations. In Fig.~\ref{widefield} we show a wide-field image in which the field of view (FOV) is 500$\times$500 mas$^2$ made from the 2001 epoch observations.  The peak flux density is 1.24\,Jy/beam and no secondary image of the core is seen down to 1.2 mJy/beam, the level of the second contour. Thus we can definitively rule out a secondary image of the core down to 0.1\% of the core brightness out to 116\,mas from the core, corresponding to an impact parameter of $5.4456 \Theta_{\rm E}$ (see Table \ref{lensvals}).  Since the smallest plausible value of the Einstein radius is $\Theta_{\rm E} $ = 1 mas, this means that the impact parameter, i.e., the angle between the jetted-AGN nucleus and the centroid of the spiral galaxy near-infrared contours, is at least 5.4456\,mas.  
   
Our conclusion is that,  in spite of the presence of a $\sim 10^8 M_\odot$ SMBH in the spiral galaxy, we see no evidence of multiple images of the compact radio core.

The absence of a secondary image might be taken as evidence that the radio source is not a background source, but is located in the spiral galaxy. But we have seen that \citet{2002AJ....124.2401P} measured an offset of $13 \pm 4$\,mas, and that the offset could be as much as 26\,mas. So the absence of a secondary image is not at all surprising and cannot be used as an argument against the background source hypothesis.
    
The lack of multiple images down to a magnification ratio of 1:1000 indicates that the displacement of the SMBH in the spiral from the radio core must be at least $(5.446 + 0.178)$\,mas = 5.624\,mas (see Table \ref{lensvals}).

\subsection{Milli-lensing by a $\sim 10^4 M_\odot$ Mass Condensate Associated with the Foreground Spiral Galaxy}\label{millilens}

\begin{figure}
    \centering
    \includegraphics[width=0.99\linewidth]{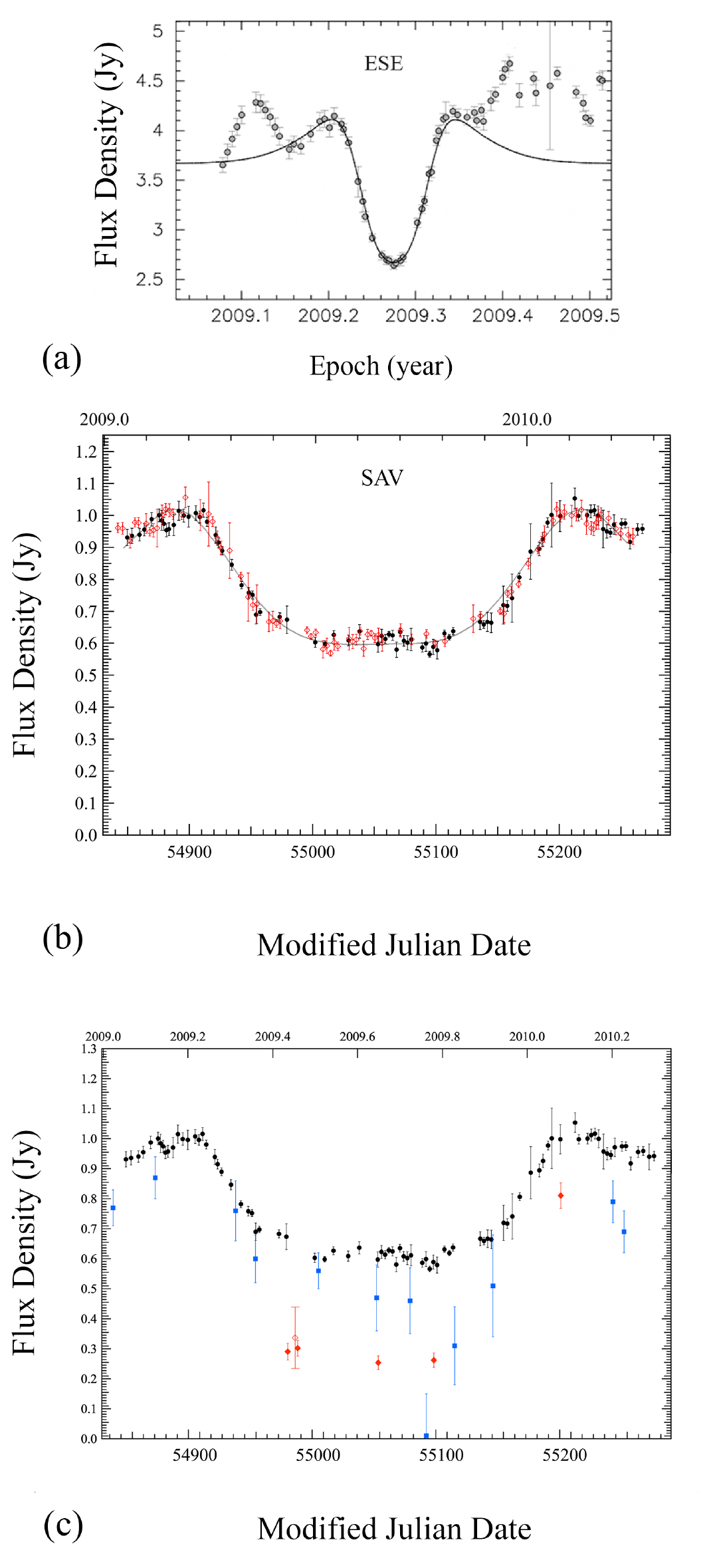}
    \caption{Symmetry in ESE and SAV. (a) Example of ESE symmetry in the blazar 2023+335 \citep{2013A&A...555A..80P} from the 40\,m Telescope monitoring program, with the black curve showing the fit of the ESE model. (b) The first SAV identified (in PKS 1413+135) on the 40\,m Telescope monitoring program reported in \citetalias{2017ApJ...845...89V}: black dots --  original data; red diamonds -  original data in reverse time order, demonstrating the high degree of symmetry. Grey curve: milli-lensing model from Paper 1.  Similarities between ESE and SAV are clear. However ESE has been definitively ruled out as the origin of SAV (see text). (c) Achromaticity of SAV from 15  GHz to 230 GHz: the black dots are from the OVRO at 15\,GHz, the blue squares are from the  Mets\"ahovi Radio Observatory at 37\,GHz, the red filled triangles are from the Submillimeter Array (SMA) at 230\,GHz, and the red open triangle is from the SMA at 345\,GHz.} 
    \label{esesav}
\end{figure}

In \citetalias{2017ApJ...845...89V} we reported the discovery of SAV events in the PKS\,1413+135 15\,GHz lightcurve, and we presented a number of arguments supporting the hypothesis that SAV is caused by gravitational milli-lensing by a $\sim 10^4 M_\odot$ mass condensate.  \citet{2017ApJ...845...90V} have shown that the SAV events are definitely not extreme scattering events \citep[ESE,][]{1987Natur.326..675F,1994ApJ...430..581F}. An example of an ESE in the blazar 2023+335 from the 40\,m Telescope 15 GHz monitoring program  is shown in Fig.~\ref{esesav}(a) \citep{2013A&A...555A..80P}. This is the highest frequency at which an ESE has been reported. Almost all ESEs detected thus far have not been visible even at 8 GHz \citep{1994ApJ...430..581F}. ESEs cannot explain SAV because that would require the angular size of the lensed component to scale as $\lambda^2$ from 15 GHz to 230 GHz, and in addition in PKS 1413+135, given that the SAV events last for one year, aberration would destroy the symmetry of any ESE.

In Fig.~\ref{esesav}(b) the original light curve is shown by the black dots, and these have been reflected about the center of the SAV event and are shown by the  red diamonds to demonstrate the very high degree of symmetry in this SAV event.  As mentioned above, and demonstrated by modelng in \citet{2017ApJ...845...90V}, this could not happen in an ESE.
 
 The putative lensed components responsible for SAV are moving at apparent speeds $\sim c$, which are typical for the apparent speeds of components in the jet of this object, as can be seen on the MOJAVE website \citep{2016AJ....152...12L,2018ApJS..234...12L}.\footnote{\url{http://www.physics.purdue.edu/astro/MOJAVE/index.html}} It is interesting to note that an intervening spiral galaxy would provide a natural place to host a milli-lens. 
 
 \subsection{Interstellar Scattering}\label{ISS}
  
It might be thought that interstellar scattering could blur any multiple images due to gravitational lensing,  and hence undermine the arguments of \S \ref{absence} and \S \ref{millilens}, but this is evidently not the case since we see from the stacked 15\,GHz images shown in Fig.~\ref{map}(d) that components of angular size $\sim 1 $\,mas are easily seen. Likewise it is very clear from the 43.2\,GHz image shown in Fig.~\ref{VLBImaps}  that components of angular size $\sim 0.5 $\,mas would easily be seen. As discussed in \citetalias{2017ApJ...845...89V} in regard to milli-lensing, non-detection of ISS at a wavelength of 18 cm \citep{1996AJ....111.1839P} shows that any scattering of the PKS 1413+135 emission by the spiral is low compared to several nuclear sightlines through our own Galaxy at low latitudes.

\section{Variability in PKS\,1413+135 from radio to $\gamma$-ray frequencies}\label{varradgamma}

We will see that a key finding of this paper is the orientation of the jet axis of the jetted-AGN PKS\,1413+135 relative to the line of sight. Our treatment relies heavily on the interpretation of the variability of this blazar, using the variability Doppler factor, $D_{\rm var}$, which is defined in Appendix~\ref{appvar}.   In Appendix~\ref{appvar} we show that $D_{\rm var}$ is a good observable in the sense that it can be determined with high accuracy from the observations. In addition to the variability Doppler factor, a number of other relationships  that are used in this section and in \S \ref{coreslm} are given, or derived, in Appendix~\ref{appvar}.

\begin{figure}
    \centering
    \includegraphics[width=0.99\linewidth]{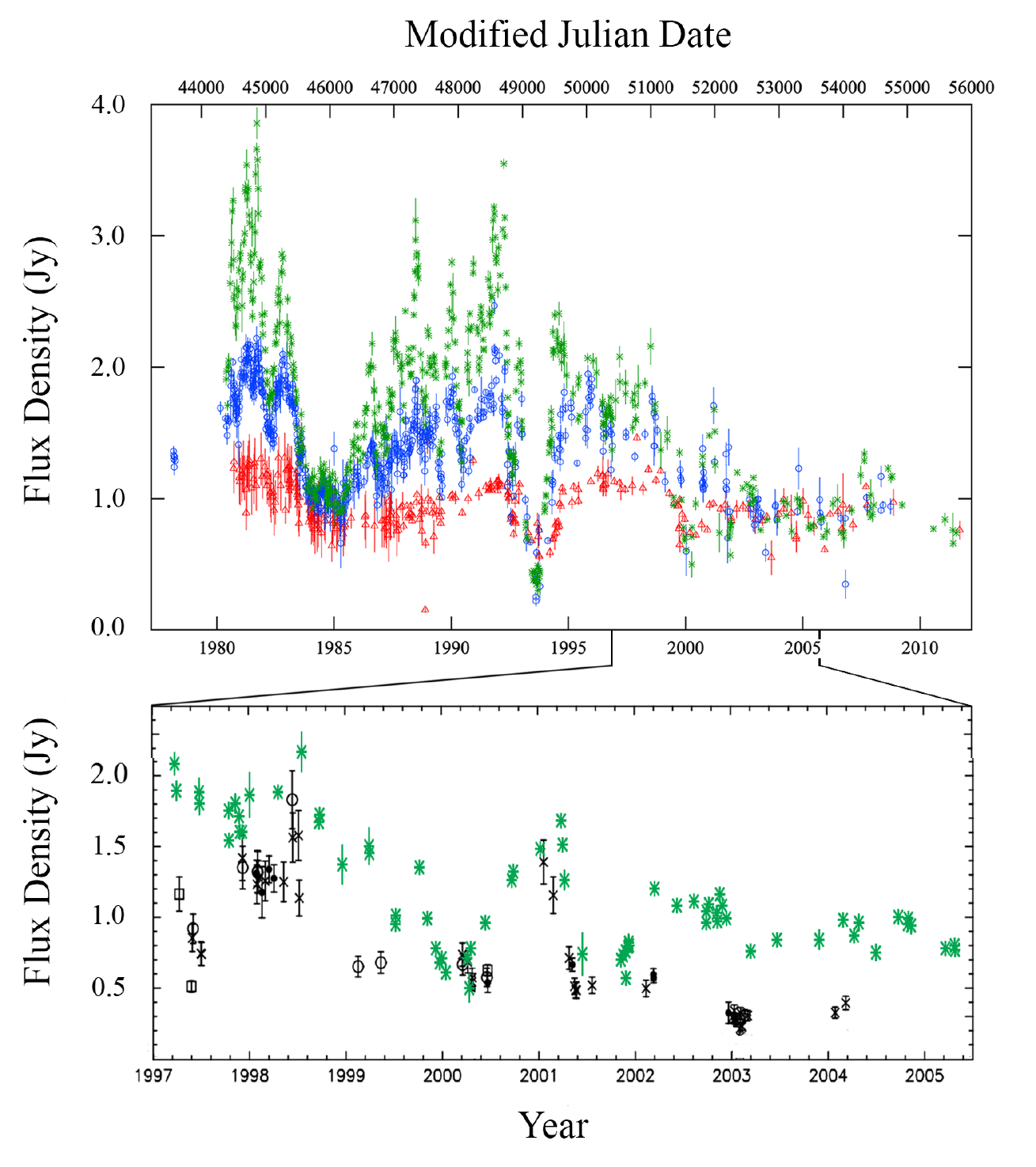}
    \caption{The radio variability of PKS\,1413+135. Upper panel: variability at radio wavelengths from UMRAO observations: red triangles - 4.8\,GHz, blue circles - 8.0\,GHz, green asterisks - 14.5\,GHz. Lower panel:  green asterisks - UMRAO 14.5\,GHz observations from the upper panel on expanded timescale; black symbols - Variability at $\lambda$850\micron\ (353\,GHz) from the JCMT \citep{2010MNRAS.401.1240J}: solid circles - three or more measurements, open circles two measurements, open boxes - two measurements with errors derived by separation of the two points, crosses - single measurements, showing primarily uncorrelated, but some correlated variations at these two frequencies (see text).}
    \label{varall}
\end{figure}

\subsection{Radio Variability in PKS\,1413+135}\label{radvar}
The radio variability of PKS\,1413+135 at 4.8, 8.0, and 14.5\,GHz from the University of Michigan Radio Astronomical Observatory (UMRAO) and at 353\,GHz from the James Clerk Maxwell Telescope (JCMT) are shown in Fig.~\ref{varall}. Examination of these light curves shows that there are some examples of strong quasi-achromatic variability, which in the most extreme cases amounts to a factor of three within a period of a few months -- see, for example the drop in flux density at both 14.5\,GHz and 353\,GHz in 2001, where the 14.5\,GHz variation mimics the 353\,GHz variation with a lag of about 70 days, similar to what was seen in \citetalias{2017ApJ...845...89V} in 2015.

There are also many examples of strong chromatic variability. For example in the first half of  1997 the 353\,GHz flux density dropped by a factor of two while the 14.5\,GHz flux density
varied little and then in the second half of 1997 the 353\,GHz flux density increased by a factor of two while the 14.5\,GHz flux density again varied little.  Another example of chromatic variability occurs between early 2002 and early 2003: here the 353\,GHz flux density drops by a factor of 2 while the 14.5\,GHz flux density drops by 20\%.
The chromatic variations in flux density are most likely to be intrinsic source variations and not due to propagation effects.  On the other hand the achromatic variations could well be due to gravitational milli-lensing and proper motion of the emission regions, as discussed in detail in \citetalias{2017ApJ...845...89V}.  
In analyzing the radio variability of PKS\,1413+135 we have therefore to be cognisant of the fact that some of the variability could be enhanced by gravitational milli-lensing.

\newpage
\subsubsection{The Radio Doppler Variability Factor of PKS\,1413+135}\label{varsav}

\citet{2018ApJ...866..137L} carried out a Doppler variability analysis of 1029 sources in the OVRO 40\,m Telescope 15\,GHz monitoring program \citep{2011ApJS..194...29R}.  The results for PKS\,1413+135 and two comparison sources, 3C\,273 and 3C\,279, are shown in Table \ref{superluminal}. The blazars 3C\,273 and 3C\,279 were chosen as comparison objects since in their high variability and luminosity they are similar to PKS 1413+135 and  also because their redshifts, $z = 0.158$ and $z = 0.536$, respectively, straddle PKS\,1413+135. \citet{2018ApJ...866..137L} assumed $z = 0.247$ for PKS\,1413+135, so we have repeated the calculations for $z = 0.5$. These results are also shown in Table~\ref{superluminal}.  In the case of PKS\,1413+135 there are two SAV events in the OVRO lightcurve (Paper 1) which we do not think are due to intrinsic fluctuations, and which might affect these values. 

In order to check for the possible effects of SAV on the determination of $D_{\rm var}$, $\gamma$, and $\theta$, we have analyzed the UMRAO 14.5\,GHz lightcurve of PKS\,1413+135 over the period 1982 December 20 to 1992 June 12 when there were no SAV events (see Fig.~\ref{varall}). These results are also shown in Table \ref{superluminal} for both $z = 0.247$ and $z = 0.5$.

\begin{deluxetable*}{l@{\hskip 8mm}ccccccc}
\tablecaption{Parameters for Blazar Jets Derived from Variability and VLBI\label{tab:}}
\tablehead{ & $z$& $D_{\rm var}$ & $\gamma$ & $\theta$ &$\theta_{\rm min}$&$\theta_{\rm max}$& $R$
}

\startdata
3C\,273& 0.158& $5.2 \pm 2.1$ & $9.4 \pm 2.8$ & $7.7\degr \pm 3.2\degr$ & &&\\
& &$3.78_{-0.55}^{+1.1}$&31.31 & $7.23\degr$&$0.36\degr$&$15.36\degr$&$1.3\times 10^7$ \\
3C\,279& 0.536&$18.3 \pm 1.9$ &$13.3 \pm 0.6$& $1.9\degr \pm 0.6\degr$ &&& \\
& &$11.64_{-1.11}^{+1.77}$&24.06 & $4.22\degr$&$0.04\degr$&$4.93\degr$&$1.8\times 10^8$ \\
PKS\,1413+135(OVRO)& 0.247&$8.59_{-1.73}^{+5.58} $ & $4.49 $ & $2.36\degr $ &$0.07\degr $&$6.68\degr $& $4.6\times 10^5$\\
PKS\,1413+135(UMRAO)&0.247&$9.63_{-2.60}^{+4.53} $ & $5.14_{-1.70}^{+1.74} $ & $1.97\degr $ & $0.03\degr$ &$5.85\degr$  & \\
PKS\,1413+135(OVRO)& 0.5&$12.19_{-1.92}^{+2.45} $&$6.44_{-0.96}^{+1.00} $ &$2.0\degr $ &$0.02\degr $ &$4.7\degr $&  \\
PKS\,1413+135(UMRAO)&0.5&$16.43_{-4.89}^{+6.85} $ & $8.58_{-2.80}^{+2.94} $ & $1.09\degr $ & $0.01\degr $ & $3.41\degr $ & \\
\enddata
\tablecomments{The variability Doppler factors, variability $\gamma$ factors, and variability angles are based on $\beta_{\rm max}$. These are calculated according to \citet{2018ApJ...866..137L} using the MOJAVE 15\,GHz $\beta_{\rm max}$ values, except for the first row on 3C\,273 and 3C\,279, which are from \citet{2017ApJ...846...98J} using their 43\,GHz $\beta_{\rm max}$ values, with 3C\,273 being revised values from Jorstad (private communication).  The UMRAO data cover a period when there are no SAV events (see text).  $R$ is the predicted jet-to-counterjet flux density ratio assuming spectral index $\alpha=0$.}
\label{superluminal}
\end{deluxetable*}
 
The values from the OVRO light curve including the two SAV events and from the UMRAO lightcurve with no SAV events are in good agreement, so the SAV events have not significantly altered the result. A key result here, as can be seen in Table~\ref{superluminal}, is the small angle $\theta$ between the jet axis and the line of sight in PKS\,1413+135.  We return to this point at the end of this section.

\begin{figure*}
    \centering
    \includegraphics[width=0.95\linewidth]{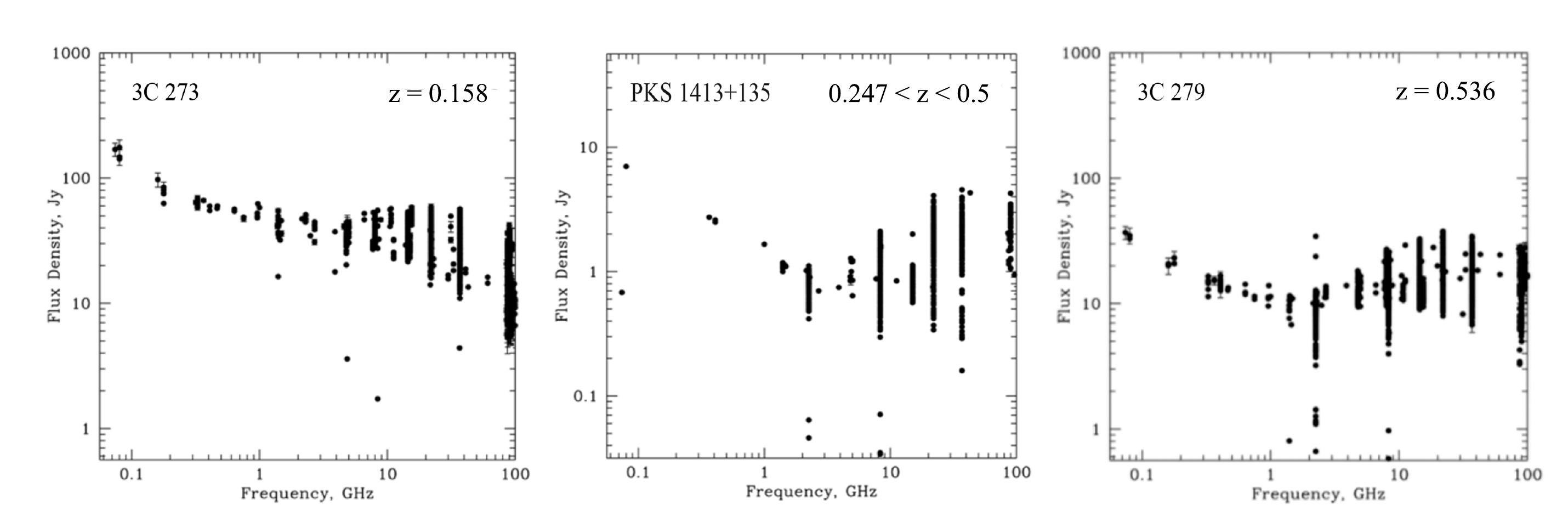}
    \caption{Jetted-AGN radio spectra and variability taken from the MOJAVE website \citep{2019ApJ...874...43L}. All of the data shown here are total flux densities, so the spread of values at each frequency is an indication of the level of variation of the flux density at that frequency over the last three decades. Shown here are the three blazars 3C\,273, 3C\,279 and PKS\,1413+135. 3C\,273 and 3C\,279 are archetypal high-luminosity blazars whose redshifts bracket that of PKS\,1413+135.}
    \label{threespectra}
\end{figure*}

\subsubsection{The Radio Spectrum of PKS\,1413+135}\label{specvar}

The radio spectra of the three high-luminosity blazars (PKS\,1413+135, 3C\,273, and 3C\,279) are shown in Fig.~\ref{threespectra}.  
All the points plotted in Fig.~\ref{threespectra} are total flux densities. The spread of points at a given frequency is an indicator of the variability since 1980 in the total flux density in these objects.  

In this paper we define the spectral index $\alpha$ by $S_{\nu}\propto \nu^\alpha$. 
At frequencies above 3\,GHz in all three sources shown in Fig.~\ref{threespectra}  the typical spectral index  is flatter than $\alpha = -0.5$ at any given epoch, and the spectrum is sometimes inverted ($\alpha>0$).
Both the spectrum and the variability of PKS\,1413+135 above 5\,GHz are very similar to those of 3C\,273 and 3C\,279, both of which have jet axes aligned close to the line of sight.

\subsection{Variability in PKS\,1413+135 at Infrared Wavelengths}\label{infrared1}

The results in Fig.~\ref{varall} show that PKS\,1413+135 is a highly variable source from radio to submillimeter wavelengths. However the first indication of such high variability in PKS\,1413+135 came from the infrared observations of \citet{1981Natur.293..714B}.  They reported that PKS\,1413+135 is ``amongst the most highly variable extragalactic sources known'' and that at 2.2\,\micron\ it had shown changes of $>20\%$ on timescales of 1\,day, and on three occasions the intensity had changed by over a factor of two in 1\,month or less.  \citet{1988AJ.....95..307I} provided further evidence of the high level of variability at mid-far infrared wavelengths in PKS\,1413+135 from {\it IRAS} observations, which showed $\sim 20\%$ variability at 60\micron\ on a timescale of 21\,days. Since the infrared emission is so highly variable it must be dominated by synchrotron radiation that is strongly beamed.
   
A question that naturally arises is whether the infrared radiation comes from the same object as the jetted-AGN radio emission. Given that we see highly variable blazar activity in both the infrared and radio wavebands, it is clear that the infrared and the radio emission both come from the same object. This is also true of the $\gamma$-ray emission.

\begin{figure}
  \centering
   \includegraphics[width=0.99\linewidth]{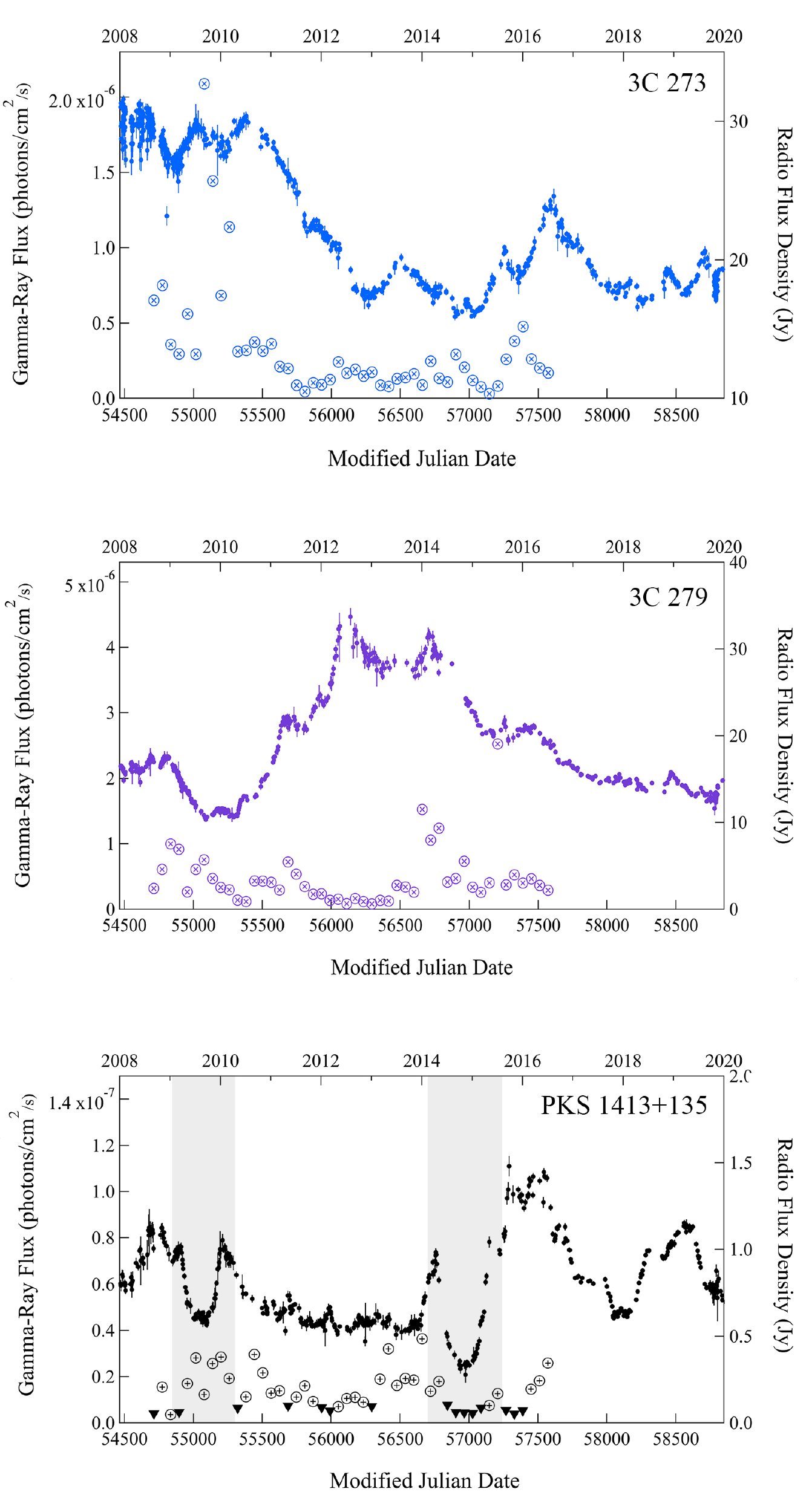}
   \caption{Radio and $\gamma$-ray lightcurves of 3C\,273, 3C\,279 and PKS 1413+135: Radio lightcurves are from the OVRO 40 m Telescope 15\,GHz monitoring program \citep{2011ApJS..194...29R} (\url{https://www.astro.caltech.edu/ovroblazars/}); $\gamma$-ray lightcurves are from the Fourth {\it Fermi\/}-LAT catalog \citep{4FGL}. The solid symbols are the radio data and the open symboes are the $\gamma$-ray data. For the $\gamma$-ray lightcurves we use the 2-month integrations at energies between 50 MeV and 100 GeV from Aug 2008 until Aug 2016. As can be seen here, PKS\,1413+135 is approximately an order of magnitude weaker than 3C\,273 and 3C\,279 in both $\gamma$-rays and at radio frequencies. On PKS\,1413+135 the triangular symbols are upper limits. On both  3C\,273 and 3C\,279 the error bars are smaller than the circular symbols. The greyed intervals in the PKS\,1413+135 lightcurve pick out the two SAV events that we ascribe to gravitational milli-lensing that were reported in \citetalias{2017ApJ...845...89V}.}
   \label{lightcurves}
\end{figure}

\subsection{Variability in PKS\,1413+135 at X-ray Energies}\label{xvar}

Based on the soft X-ray absorbing column density, \citet{1992ApJ...400L..17S} estimate that $A_{V}\gtrsim30$\,mag of extinction towards the X-ray emission regions in the PKS\,1413+135 AGN and they estimate a column density $N({\rm H})\ge 2 \times 10^{22} \; {\rm cm^2}$ along the line of sight to this source. Because it is so heavily absorbed at soft X-ray energies, \citet{2002AJ....124.2401P} observed PKS\,1413+135 over the 2--10\,keV energy range with \textit{ASCA} and obtained a detectable spectrum.  They derived a luminosity  of $P(2$--$10\,{\rm keV}) = 2.2 \times 10^{44}$\,erg\,s$^{-1}$.  This is a factor $\sim 5$ lower than in earlier X-ray observations,  so it is clear that the jetted-AGN PKS\,1413+135 is highly variable at X-ray energies.

\subsection{Variability in PKS\,1413+135 at $\gamma$-ray Energies}\label{gammavar}

 The $\gamma$-ray and radio emission from blazars has been compared in a number of studies \citep[e.g.,][]{2010ApJ...718..587M,2011A&A...532A.146L,2015ApJ...810L...9L,2015MNRAS.452.1280R,2020MNRAS.492.3829L}. In particular it is well known that {\it Fermi}-LAT preferentially detects highly Doppler-boosted jets \citep{2009ApJ...696L..17K,2010ApJ...715..429A,2010A&A...512A..24S,2014MNRAS.438.3058R}.
\citet{2015ApJ...810L...9L} show that in blazars there is a strong correlation between Doppler factor and $\gamma$-ray flux.

The SED of PKS\,1413+135 peaks at $10^{13.4}$\,Hz, then drops precipitously \citep{2011A&A...536A..15P}, but this is strongly affected by extinction, so it is likely that the true peak lies at  higher frequency. In this regard PKS\,1413+135 is typical of $\gamma$-ray blazars and of blazars with high Doppler factors in general \citep{2015ApJ...810L...9L}.  

Lightcurves at 15\,GHz from the OVRO 40\,m Telescope and at $\gamma$-ray energies from the fourth Fermi-LAT catalog \citep{4FGL} for 3C\,273, PKS\,1413+135, and 3C\,279 are shown
 Fig.~\ref{lightcurves}.  Note that the ratio of the $\gamma$-ray flux to the radio flux density is very similar in all three objects. In addition, all three objects show very similar variability to each other at both radio frequencies and at $\gamma$-ray energies.

Recently a burst of $\gamma$-ray emission from PKS\,1413+135 was detected with {\it Fermi}-LAT \citep{2019ATel13049....1A}: on 2019 August 28 the daily averaged flux at $E>100$ MeV reached  $(4.8 \pm 1.5) \times 10^{-7}$ photon\,cm$^{-2}$\,s$^{-1}$, about 38 times the average flux reported in the fourth \textit{Fermi}-LAT catalog. During the outburst the $\gamma$-ray spectrum was significantly harder than the average from this source. This hard-spectrum state was accompanied by the detection of several $E>10$\,GeV photons, including one with an energy of $\sim 64$\,GeV.

\subsection{The Variability of PKS\,1413+135: Conclusions}\label{concvar}

We see no other way to explain the observed strong variability of PKS\,1413+135 across the electromagnetic spectrum, from radio frequencies to $\gamma$-ray energies than by relativistic beaming, and we find this a compelling demonstration of strong Doppler boosting as expected in a jet close to the line of sight.  We return to this in \S \ref{coreslm}, where we analyze the subluminal and superluminal motion of PKS\,1413+135 and show it to be entirely consistent with our findings based on variability.  Our key conclusion of this section is that, as illustrated in Fig.~\ref{cartoon},  the jet axis in PKS\,1413+135 is aligned to between $0.01\degr$ and $6.68\degr$ of the line of sight taking the most conservative values we present in Table \ref{superluminal} from the combination of the OVRO and UMRAO results. This value is refined in \S \ref{coreslm} based on the MOJAVE observations of $\beta_{\rm app}$.

 \section{Superluminal Motion in PKS\,1413+135}\label{coreslm}
    
Having determined that PKS\,1413+135 lies in the redshift range $0.247< z < 0.5$ (see \S \ref{bulge}), we can interpret the proper motion of components in the PKS\,1413+135 jet and counterjet. Hence we can refine our determination of the orientation of the radio jet axis in PKS\,1413+135 relative to the line of sight. The reader is referred to Appendix~\ref{appvar} for parameter relationships used in this section.

\subsection{Subluminal and Superluminal Motion in PKS\,1413+135}\label{beaming}

The total flux density of most blazars at 15\,GHz is dominated by the core flux density, which is strongly relativistically beamed. The cores of blazars are generally stationary features, which provide a convenient reference point against which the relative motion of components moving out from the core along the jet or counterjet can be measured.  

The radio emission from PKS\,1413+135 at 8\,GHz and higher frequencies is totally dominated by the emission from the core, as can be seen in Fig.~\ref{VLBImaps}.  So in this object it is a simple matter to measure the apparent motions of features in the jet relative to the core at frequencies  above 8\,GHz.  We will focus on the 15\,GHz MOJAVE observations.

The apparent transverse speed of a relativistic object is given by equation~(\ref{vapp}) in Appendix~\ref{appvar}.
We note that the highest superluminal motion observed in PKS\,1413+135 ($1.72 \pm 0.11 c$ for $z=0.247$ and a factor of 1.58 higher for $z=0.5$) is much lower than those of  3C\,273 and 3C\,279 given in the MOJAVE data base, which are  $(20.50 \pm 0.82)c$ and $(10.0\pm 0.86)c$, respectively.

\begin{figure}
   \centering
    \includegraphics[width=0.98\linewidth]{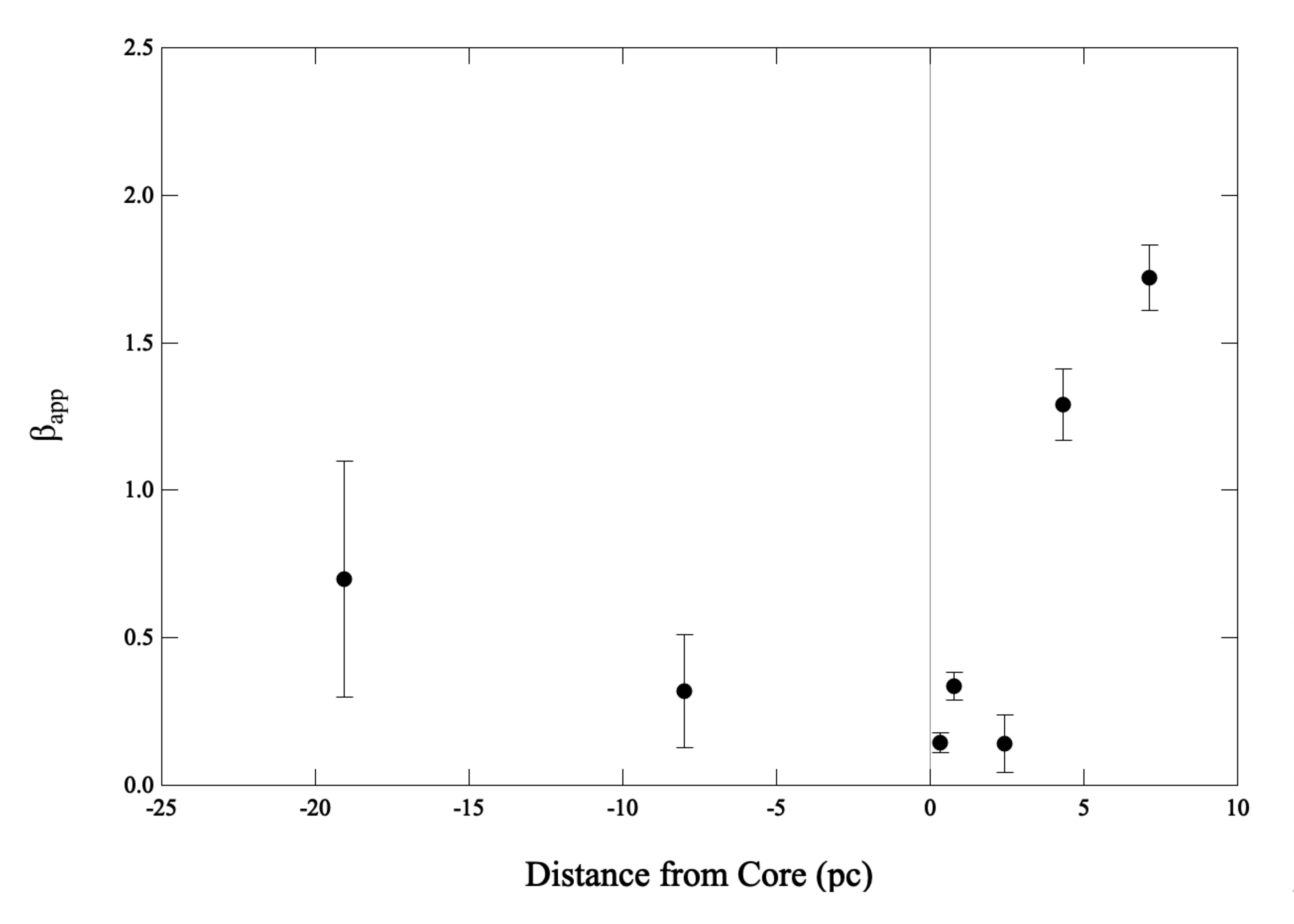}
    \caption{PKS\,1413+135 component speeds as a function of distance from the core as measured on the MOJAVE program assuming $z=0.247$. The speeds corresponding to $z=0.5$ are a factor 1.58 greater than shown here (see text).
    The vertical grey line indicates the position of the core. Negative distances from the core, i.e. points to the left of the grey line, correspond to components in the counterjet. Positive distances from the core, i.e., points to the right of the grey line, correspond to components in the jet.}
    \label{speeds}
\end{figure}

The apparent velocities, $\beta_{\rm app} = v_{\rm app}/c$, that have been measured on the MOJAVE program for seven components in PKS\,1413+135 are shown in Fig.~\ref{speeds} for five components in the  jet and two components in the counterjet. 

\subsubsection{Implications of $\beta_{\rm app}$ on the Counterjet Side}\label{counterjet}

As we can see from Fig.~\ref{speeds} it appears as if $v_{\rm app}$ for the components on the counterjet side is increasing with distance from the core, although the significance of both measurements is less than 2$\sigma$, so more observations are needed to confirm these apparent speeds.  However if this apparent increase in speeds were to be confirmed then it could not be due to relativistic effects because the apparent speed in the counterjet is approaching the velocity of light, which can only happen as $\theta \rightarrow 90\degr$, which is definitely not the case.  Thus,  were this result to be confirmed, it would indicate that we are observing a ``pattern'' speed and not the bulk motion of the emitting plasma \citep{1985ApJ...295..358L,2007ApJ...658..232C}.

\subsubsection{Implications of $\beta_{\rm app}$ on the Jet Side}\label{jet}

It can be seen in Fig.~\ref{speeds} that there is a highly significant increase in $v_{\rm app}$ on the jet side as components move away from the core. This could be due to (i) a real acceleration in these components as they move away from the core, (ii) a change in angle between the velocity vector of the component and the line of sight, or (iii) a pattern speed not related to the bulk motion of the emitting plasma. 

Since we know that the jet is bending, the simplest model is (ii), i.e., a model in which the material is moving at constant speed as it moves away from the core while the angle is changing. We have determined that $\theta <  6.68\degr$ (\S \ref{varradgamma}), so $\theta <  1/\gamma$ (see Table \ref{superluminal}). In this case, equation~(\ref{vapp}) shows that, in order to explain the increase in apparent speed with component distance from the core, the angle between the jet axis and the line of sight must be lower at the core than it is out along the jet, i.e., the jet must curve away from the line of sight as we move out along the jet.  

 Note that it must be the case that the bulk velocity of the plasma emerging from the core on the jet side is highly relativistic, otherwise the high variability of the core 
 %of PKS\,1413+135 
 could not be explained by relativistic beaming.

In Table \ref{angles} we list the angles at which the  speeds $\beta=0.13c$ and $\beta=1.72c$ (the range of speeds seen in the jet in Fig.~\ref{speeds} assuming $z=0.247$) would be observed for different values of $\gamma$. We also show the values for speeds a factor 1.58 greater, corresponding to $z=0.5$. We also list the Doppler factors, $D_0$, for $\theta=0\degr$, and the maximum angle $\theta_{\times 100}$ at which relativistic beaming would boost the observed flux density by at least  a factor 100.

For example, if $\gamma = 5$ then $\beta_{\rm app} \sim 0.13$ requires $\theta \sim 0.15\degr$ at a projected distance of 0.3\,pc from the core (as shown in Fig.~\ref{speeds}),  and $\beta_{\rm app} \sim 1.72$ requires $\theta \sim 2.1\degr$ at a projected distance of 7\,pc from the core (as shown in Fig.~\ref{speeds}); while for
$\gamma = 4$ we have that $\beta_{\rm app} \sim 0.13$ requires $\theta \sim 0.25\degr$ at a projected distance of 0.3\,pc from the core, and $\beta_{\rm app} \sim 1.72$ requires $\theta \sim 3.4\degr$ at a projected distance of 7\,pc from the core.

{\catcode`\&=11
\gdef\2009AandA...494..527H{\citet{2009A&A...494..527H}}}

\begin{deluxetable}{l@{\hskip 8mm}cccccc}
\tablecaption{Viewing Angle Constraints from $v_{\rm app}$\label{tab:xyz1}}
\tablehead{ $\gamma$& $D_0$ & $\theta_{0.13c}$ & $\theta_{1.72c}$ & $\theta_{0.20c}$ & $\theta_{2.71c}$& $\theta_{\times 100}$\\
&&$z=0.247$&0.247&0.5&0.5&}

\startdata
2.2  & 4.16 & $0.9\degr$ &$16.2\degr$ &$1.4\degr$&- &$3.3\degr$ \\ 
2.5  & 4.79 & $0.7\degr$ & $10.8\degr$ &$1.05\degr$&- &$10.0\degr$\\
3  & 5.83 & $0.45\degr$ &$6.7\degr$ &$0.7\degr$&$14.6\degr$ &$13.0\degr$ \\ 
4  & 7.87& $0.25\degr$ &$3.4\degr$ &$0.38\degr$&$5.935\degr$ &$14.1\degr$ \\ 
5  & 9.90 & $0.15\degr$ &$2.1\degr$ &$0.235\degr$&$3.49\degr$ &$13.86\degr$ \\ 
6  & 11.92 & $0.11\degr$ &$1.42\degr$ &$0.163\degr$&$2.33\degr$ &$13.35\degr$ \\ 
7  & 13.93 & $0.075\degr$ &$1.05\degr$ &$0.12\degr$&$1.675\degr$ &$12.79\degr$ \\ 
\enddata
\tablecomments{The viewing angles, $\theta$, at which the apparent speeds are $v_{\rm app} = 0.13 c$ and $v_{\rm app} = 1.72 c$, corresponding to a redshift of 0.247, and $v_{\rm app} = 0.20 c$ and $v_{\rm app} = 2.71 c$, corresponding to a redshift of 0.5. $D_0$ is  the on-axis Doppler factor, $D(\theta=0)$. The range of viewing angles over which Doppler boosting in the approaching jet exceeds a factor 100 is from $\theta = 0\degr$ to $\theta = \theta_{\times 100}$.}
\label{angles}
\end{deluxetable}

For illustrative purposes we assume  two Lorentz factors for the bulk motion along the jet: $\gamma = 7$ and $\gamma = 3$. From equation~(\ref{vapp}) we find that the observed maximum apparent speed of $1.72\, c$ implies $\theta \sim 59\degr$ or $\sim 1\degr$ for $\gamma=7$, and  $\theta \sim 54\degr$ or $\sim7\degr$ for $\gamma=3$. We have shown in \S \ref{varradgamma} that the jet axis in PKS\,1413+135 is aligned close to the line of sight so we can rule out the larger option, $\theta \sim 59\degr$ or $\sim 54\degr$.

The solution  $\gamma =7$ is fully consistent with the requirements of beaming, but $\gamma =3$ is more likely since the required alignment of the innermost features is less stringent. Note that for redshifts of the radio source close to $z = 0.247$ $\gamma=2.5$ is likewise entirely feasible, and that $\gamma=2.2$ also works (just); whereas for redshifts of the radio source close to $z = 0.5$ $\gamma=4$ is entirely feasible, and  $\gamma=3$ also works (just).

Inspection of the angles $\theta$ given in Table \ref{angles} at which the apparent speeds could be observed shows that we require $\theta < 1\degr$ to account for the very low apparent speeds measured within the first few parsecs of the core shown in Fig.~\ref{speeds}.
    
    Our conclusions of this section are thus that the PKS\,1413+135 jet axis close to the core is inclined at an angle $\theta < 1\degr$ to the line of sight, and bends away from this angle as we move out along the jet, and that the high observed flux density,  variability, and the apparent speeds, are easily explained by relativistic beaming in terms of the entirely self-consistent model suggested above.  Furthermore, the values of the key parameters in PKS\,1413+135 are very similar to those of 3C\,273 and 3C\,279, shown in Table \ref{superluminal}.

\section{The Location of PKS\,1413+135 Relative to the Spiral Galaxy}\label{assoc}

We now consider the possibility that the spiral galaxy is the host of the jetted-AGN blazar PKS\,1413+135.

There are two arguments that taken together make an extremely strong, if not absolutely watertight, case that the jetted-AGN PKS\,1413+135 is not located in the spiral galaxy. These are based on (i) the absence of continuum emission signatures of re-processed optical-ultraviolet radiation in the infrared spectrum of the spiral galaxy, and (ii) the fact that the angular momentum axis of the jetted-AGN central engine is orthogonal to that of the spiral galaxy. We discuss these two arguments below.

\subsection{The Bright Infrared Nucleus and the Infrared Spectrum}\label{infrared2}

If the AGN PKS\,1413+135 is located in the spiral galaxy, much of the  optical-ultraviolet radiation from AGN will be absorbed in the galactic disk and re-radiated at infrared wavelengths.  We therefore first consider the luminosity of the AGN in the optical--ultraviolet frequency range.

\subsubsection{The Luminosity of the Blazar PKS\,1413+135}\label{optsp}

 We cannot determine the flux density or variability of the jetted-AGN PKS\,1413+135 at optical and ultraviolet wavelengths from direct observations in these energy bands because the blazar  optical and ultraviolet continuum is totally obscured by the high extinction arising in the edge-on spiral galaxy. 
\citet{2011A&A...536A..15P} show that in PKS\,1413+135 the SED ($\nu F_{\nu}$)  falls by an order of magnitude between $10^{13.5}$\,Hz and $10^{14}$\,Hz, i.e., a drop of 1.5 orders of magnitude in $F_{\nu}$ over this frequency range. The SED  then drops another two orders of magnitude by $10^{15}$\,Hz. The infrared spectrum yields an $A_{V}$ of 14\,mag (see  \S \ref{infraredsp} below) and the X-ray spectrum yields an $A_{V}$ of 30\,mag (\S \ref{xvar}). Thus it is  not possible to obtain direct measurements of the SED of the jetted-AGN PKS\,1413+135 in this critical wavelength range. We therefore have to estimate this by comparison with other blazars.  

From \citet{2002ApJ...567...50B}, \citet{2006A&A...451L...1T}, and  \citet{2011A&A...536A..15P}, we find the SEDs of 3C\,273 and 3C\,279 in the optical--ultraviolet bands that are obscured in PKS\,1413+135.  We obtain estimates of the likely luminosity of the PKS\,1413+135 jetted-AGN in the two cases: (i)~assuming that its SED shape is similar to 3C\,273, and (ii)~assuming its SED shape is similar to 3C\,279. 
Since we wish to calculate the effect of the jetted-AGN PKS\,1413+135 if it is located in the spiral galaxy, we compare  its luminosity to those for 3C\,273 and 3C\,279 over the same rest-frame frequency range. The plots of \citet{2011A&A...536A..15P} show that the SED ($\nu F_\nu$) of 3C\,273 is flat between $10^{14}$\,Hz and $10^{15.3}$\,Hz, 
and the  SED of 3C\,279 drops by a factor of 20 over this frequency range. \citet{2006A&A...451L...1T} found that the SED of 3C\,273 rose by a factor of three between $10^{14}$\,Hz and $10^{15.3}$\,Hz, and \citet{2002ApJ...567...50B} found that the SED of 3C\,279 fell by a factor of $14.7 \pm 3.3$ between the peak at $10^{13}$\,Hz and $10^{15.3}$\,Hz, where the uncertainty is due to variability not measurement errors. In our modeling of PKS\,1413+135  we therefore conservatively assume a flat SED in the case of 3C\,273, and a drop of a factor of 20 in the SED of 3C\,279.

The corresponding luminosities, and their ranges due to variability, that we derive  are given in Table \ref{lums}. As a consistency check we note that these luminosities are in good agreement with those of \citet{1964ApJ...140....1G} for 3C\,273, and \citet{1990AJ....100.1452W} for 3C\,279.

We obtain a lower limit to the luminosity of the jetted-AGN PKS\,1413+135 by assuming that it is at a redshift of $z = 0.247$. Under this assumption we find that the luminosity of PKS\,1413+135 is 42\% of the luminosity of 3C\,273 if it has the same SED shape as 3C\,273 and 9\% of the luminosity of 3C\,279 if it has the same SED shape as 3C\,279.  The luminosities of 3C\,273 and 3C\,279 over this range differ by only a factor $\sim 3$.  So our conclusion is that the  optical--ultraviolet luminosity of the jetted-AGN PKS\,1413+135 lies in the range $9\%$--$42\%$ of the luminosities of 3C\,273 and 3C\,279 on the assumption that the shape of its SED lies in the range bracketed by the SED shapes of 3C\,273 and 3C\,279.  We see from Table \ref{lums} that, taking the more conservative value given by a drop in the SED by a factor of 20 between $10^{14}$\,Hz and $10^{15.3}$\,Hz, the mean luminosity  of PKS\,1413+135 over the range $3.3 \times 10^{14}$\,Hz to $3.3 \times 10^{15}$\,Hz is $10^{45} \; {\rm erg \, s^{-1}}$.

\begin{deluxetable}{l@{\hskip 8mm}ccc}
\tablecaption{Luminosities\label{tab:xyz11}}
\tablehead{ Blazar& $\delta {\rm SED}$&Luminosity&Range\\
& & ${\rm erg \; s^{-1}}$& ${\rm erg \; s^{-1}}$}
%\decimalcolnumbers
\startdata
3C\,273  &1 &$4.6 \times 10^{46}$&$2 \times 10^{46}$--$9 \times 10^{46}$ \\ 
3C\,279 & 20 &$1.2 \times 10^{46}$&$4 \times 10^{45}$--$4 \times 10^{46}$\\ 
PKS\,1413+135&1 &$2 \times 10^{46}$& $10^{46}$--$4 \times 10^{46}$\\
PKS\,1413+135&20 &$ 10^{45}$& $3 \times 10^{44}$--$3 \times 10^{45}$\\
\enddata
\tablecomments{Luminosities over the frequency range $3.3 \times 10^{14}$\,Hz to $3.3 \times 10^{15}$\,Hz (the Lyman limit). The factor $\delta {\rm SED}$ is the factor by which the SED, $\nu F_\nu$, drops between the observed SED peak in the infrared and the SED value in the optical band for 3C\,273 and 3C\,279. The luminosity range accounts for variability in 3C\,273 and 3C\,279. The values for PKS\,1413+135 are two values assumed based on the shapes of the SEDs of 3C\,273 and 3C\,279, and under the assumption that the redshift of the blazar PKS\,1413+135 is $z = 0.247$, i.e., that it is located in the spiral galaxy (see \S\ref{optsp}).}
\label{lums}
\end{deluxetable}

\begin{figure}
    \centering
    \includegraphics[width=0.95\linewidth]{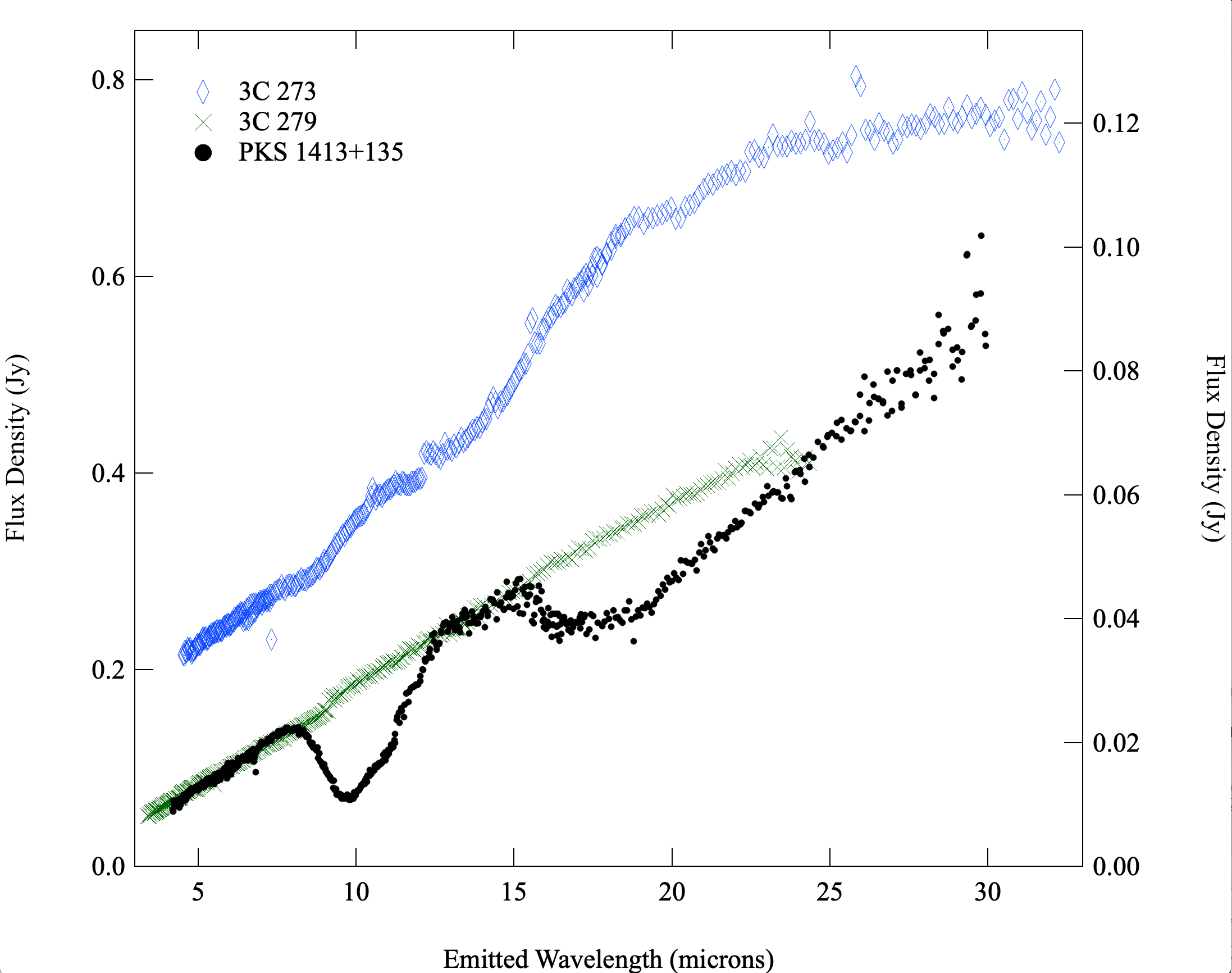}
    \caption{\textit{Spitzer} spectra of 3C\,273 (blue diamonds), 3C\,279 (green crosses) and PKS\,1413+135 (black dots).The flux densities of 3C\,273 and 3C\,279 are given on the left axis. The flux density of PKS\,1413+135 is given on the right axis. 3C\,273 shows a significant bump of emission in the continuum and the two silicate emission features at rest-frame 9.7\,\micron\ and 18\,\micron\ are clear. The scale of the right axis has been adjusted such that the PKS\,1413+135  spectrum fits the 3C\,279 spectrum outside of the silicon absorption features.}
    \label{irspectra}
\end{figure}

\subsubsection{The Mid-Infrared Spectrum of PKS\,1413+135}\label{infraredsp}

The strong variability of PKS\,1413+135 at infrared wavelengths discussed in \ref{infrared1} shows that it is dominated by strongly beamed synchrotron radiation.
In Fig.~\ref{irspectra} we show \textit{Spitzer} spectra for PKS\,1413+135, 3C\,273, and 3C\,279 \citep{2004ApJS..154....1W,2004ApJS..154...18H}.  We present a re-reduction of the PKS\,1413+135 spectrum, using the Cornell Atlas of \textit{Spitzer}/IRS Sources (CASSIS) \citep{2011ApJS..196....8L}, of the \textit{Spitzer} low-resolution spectrum  previously published by \citet{2010ApJ...713.1393W}, together with \textit{Spitzer} CASSIS spectra of 3C\,273 and 3C\,279. We used CASSIS version~7, and the AORKEYs as follows: PKS\,1413+135: 18322432, 3C\,273: 28976640, and 3C\,279: 27437312.  It can be seen here that  3C\,273 shows a mid-infrared bump of emission in the 9.7\,\micron\ and 18\,\micron\ silicate bands \citep{2005ApJ...625L..75H}. This is due to thermal radiation arising from re-processed optical--ultraviolet emission from the AGN \citep{2007ApJ...656..148A}.

 Millimeter-wavelength observations of PKS\,1413+135 show  CO absorption of the millimeter continuum \citep{1994A&A...286L...9W}, so the continuum  at millimeter wavelengths is being viewed through a  molecular screen. The \textit{Spitzer} IRS emission-frame spectrum, assuming $z = 0.247$, is shown in Fig.~\ref{irspectra}.  This shows silicate absorption at 9.7\,\micron\ and 18\,\micron. The spectrum can be fitted by a power-law continuum absorbed by an intervening screen of cold dust, which is consistent with the CO observations, i.e., there is no evidence of significant heating of the dust.  The strength of the rest-frame 9.7\,\micron\ silicate absorption corresponds to an $A_{V}$ of  13.5--14\,mag for a standard galactic extinction law \citep{1985ApJ...288..618R} and fits the  sharp drop in flux density at near-infrared wavelengths.  A similar conclusion was reached by \citet{2010ApJ...713.1393W} based on an earlier reduction of the \textit{Spitzer} spectrum.  
 
 As shown in \S \ref{optsp}, the integrated luminosity of the jetted-AGN PKS\,1413+135 over the range $10^{13.5}$--$10^{14}$\,Hz is $\sim 10^{45}\;{\rm erg \, s^{-1}}$.
 If PKS\,1413+135 were located in the spiral galaxy, the ionizing continuum optical--ultraviolet radiation from the nucleus would be absorbed and reprocessed in the disk of the galaxy by the obscuring dust and we would expect to see a near-infrared dust continuum bump and near-infrared emission lines from dust. However one does not always see any reprocessed radiation in the form of line emission in such cases.  For example,   the ultra-luminous infrared galaxy (ULIRG) 08572+3915 shows no lines at the mid-infrared wavelengths of \textit{Spitzer} \citep{2007ApJ...656..148A}.
 
 Nevertheless, we would still expect to see evidence of a bump of emission, or unusual relative depths of the silicate absorption features at 9.7\,\micron\ and 18\,\micron\ in the \textit{Spitzer} spectrum if a powerful AGN was embedded in the spiral galaxy. However the relative depths of the two silicate absorption features are normal for absorption from cold dust.  This is strong evidence that a significant source of power, such as a blazar, is not embedded in the obscuring material in the nucleus of the spiral galaxy.

\subsection{Spin Axes of Jetted-AGNs in Spirals}
\label{misaligned}

We might expect that any central engine generating a jetted radio source in the nucleus of a spiral galaxy must have derived its angular momentum from the disk of the spiral galaxy, and so, absent a strong interaction via a merger event that torqued the spin axis of the central engine into the galactic disk plane, it would be somewhat aligned with, and not orthogonal to, the spin axis of the spiral host.  This would argue that the radio source PKS 1413+135 is not located in the spiral galaxy.

However, we have seen that the spiral is a Seyfert galaxy, and a
 number of studies have been carried out in Seyfert galaxies of the alignment of the radio jet axis with the spin axis of the spiral host galaxy, and in general Seyfert galaxies show  no such alignment of the galaxy spin axis with the jet axis \citep{1981ApJ...247..419U,1984ApJ...285..439U,1998ApJ...495..189C,1999ApJ...524..684G,2000ApJ...537..152K,2006AJ....132..546G}. Indeed in one such case (NGC 2110) the $\sim800$ kpc jet ``appears to propagate directly into the disk of the surrounding host galaxy'' \citep{2006AJ....132..546G}. \citet{2012MNRAS.425.1121H} discuss the physical processes, demonstrated with simulations, that give rise to such misalignments. 
 
 These are lower luminosity radio sources than PKS 1413+135, so the comparison may be misleading. Indeed the study of Seyfert galaxies by \citet{2006AJ....132..546G} focuses entirely on ``radio quiet'', albeit clearly not ``radio silent'', Seyfert galaxies. A clue to what may be going on comes from ${\rm H_2 O}$ mega-maser galaxies, which we discuss next.
 
\subsubsection{Alignment of Jet Axes and Accretion Disk Axes
in ${\rm H_2 O}$ Mega-Maser Galaxies}
 
 The study of the alignment of jet axes with the spin axes of the accretion disks in nearby edge-on spiral low luminosity active galaxies by \citet{2019A&A...624A..42K} provides an interesting insight into what might be going on. They observed 18 nearby low luminosity active galactic nuclei with VLBI and detected 5 of them at $8\sigma$ or higher signal-to-noise ratio. Of these five objects, four have a maser disk with known orientation, and all four jet axes lie in a cone within $32^\circ$ of the normal to the maser disk, and the misalignment is smaller when the inner radius of the accretion disk is larger. This suggests that there may be a correlation of jet/disk alignment with luminosity, since  the inner radii of the accretion disks are likely to  be larger in higher-luminosity AGN.

\subsubsection{Alignment of Jet Axes and Accretion Disk Axes
in high-luminosity AGN}

Powerful relativistic jets are rare in spiral galaxies, and we know of only eight clear examples — 3C 120 \citep{1967PASP...79..369S,2005ApJ...620..646H}, 0313$-$192 \citep{1998ApJ...495..227L}, J0836+0532, J1159+5820, J1352+3126, and J1649+2635 \citep{2015MNRAS.454.1556S},  J2345-0449 \citep{2014ApJ...788..174B},  and Speca \citep{2011MNRAS.417L..36H}. Of these J0836+0532, J1159+5820, J1649+2635 and 3C 120 are almost face-on and so it is not possible to determine the projected angle between the jet axis and the spin axis of the galactic disk, but 0313$-$192, J1352+3126, J2345$-$0449 and Speca are all highly enough inclined to the line
of sight for the projected angles between the spiral axis
and the jet axis to be determined. We have measured the position angles of the galactic disks and the radio jets by inspection of the images and maps given in the above papers, with the results shown in Table \ref{dpa}. We note that there are significant, but small,  mis-alignments in three cases (0313$-$192, J1352+3127 and Speca), but that the radio jet in J2345$-$0449 is well aligned with the normal to the galaxy disk, and in all four cases the sky-projected jet
and spiral spin axes are aligned to within closer than $25^\circ$, with three of them being aligned to closer than $10^\circ$. The Fanaroff \& Riley classes \citep{1974MNRAS.167P..31F} of these objects are also shown in Table \ref{dpa}. We note that  J2345$-$0449 has a SMBH of mass $> 2 \times 10^8 M_\odot$ \citep{2014ApJ...788..174B}. 

We have seen that PKS 1413+135 is a powerful jetted-AGN. It is therefore far more likely to be similar to the AGN of Table \ref{dpa} than to standard Seyfert galaxies or indeed to ``radio quiet'' AGN. Thus, for reasons both of its jet orientation and its luminosity we think it extremely unlikely that PKS 1413+135 is located in the Seyfert 2 spiral galaxy.

\begin{deluxetable*}{l@{\hskip 8mm}cccccc}
\tablecaption{Luminosity and Jet Axis Orientation Relative to Spiral Disk Plane \label{tab:xyz88}}
\tablehead{ Source&redshift&$P_{1.4 {\rm GHz}} $&FR&PA&PA&$|\delta PA|$\\
&&${\rm W\;Hz^{-1}}$&&galactic disk & radio jet& $|{\rm disk-jet}|$\\
&&&&(degrees) & (degrees)&(degrees)}

%\decimalcolnumbers
\startdata
0313-192 (Abell 428)&0.067 &$1.0\times 10^{24}$&I&$112.8\pm 0.2$ &$17.5\pm 1.3$&$95.3\pm 1.3$\\ 
J1352+3127&0.045&$2.2\times 10^{25}$ &II&$64.7\pm 0.5$ &$132.4\pm 1.7$&$67.7\pm 1.8$\\
J1409-0302 (Speca)&0.1378&$2.3 \times 10^{24}$&II&$19.4\pm 0.8$ &$117.0\pm 1.6$&$97.6\pm 1.8$\\
J2345-0449&0.0755&$4.9 \times 10^{24}$&II&$96.7\pm 0.4$ &$7.6\pm 0.4$&$89.1\pm 0.6$\\
\enddata
\tablecomments{For PKS 1413+135 (z=$0.247\rightarrow 0.5$), $P_{1.67 {\rm GHz}} $ = $1 \rightarrow 5 \times 10^{26}\; {\rm W\;Hz^{-1}}$ assuming that the 60\% of the 1.67 GHz flux density that comes from components A and B is isotropic (see text) }
\label{dpa}
\end{deluxetable*}

\subsection{Positional Alignment of the Blazar nucleus with the Spiral Nucleus}\label{aposteriori}

 There is one argument against this interpretation that should be mentioned here, namely the small angle projected on the sky of $13 \pm 4$\,mas ($52\pm 16$ pc) between the infrared blazar position and the infrared isophotal center of the spiral galaxy \citep{2002AJ....124.2401P}. 

Following a line of argument similar to that laid out in   \citetalias{2017ApJ...845...89V}, Appendix D, we calculate the {\it a posteriori\/} probability of this alignment.  It is {\it a posteriori\/} because we would not have thought of asking the question {\it a priori\/}, i.e. before noticing that PKS 1413+135 is closely aligned with the spiral galaxy. So we pre-selected the source before calculating the probability.

We saw in \S \ref{SMBHmass} that the  disk and bulge R magnitudes of the spiral are $m_{\rm R}^{\rm disk} = 18.9$, and $m_{\rm R}^{\rm bulge} = 20.6$ \citep{1991MNRAS.249..742M}, which yield a combined magnitude of $m_{\rm  R}^ {\rm disk+bulge} = 18.7$.
We now calculate the probability of chance alignment of a background source and a foreground galaxy of magnitude $m_{\rm  R} = 18.7$ to within an angle $\zeta_{\rm off}$. A compilation of differential number counts of galaxies in the R-band measured in 23 separate analyses of different areas of sky  is given in Fig. 12 of  \citet{2001MNRAS.323..795M}.  Integrating the differential counts down to $m_{\rm  R} = 18.7$ we find there are 631 galaxies per square degree brighter than this limit. Thus the total number of   galaxies in the sky brighter than $m_{\rm  R} = 18.7$ is $N_{\rm f}=2.60\times 10^7$.

The fraction of sky within a radial distance $\zeta_{\rm off}$ from any of the $N_{\rm f}$ foreground galaxies is $\pi\zeta_{\rm off}^2N_{\rm f}/4\pi$. The expected number of chance-alignments is therefore $N_{\rm f}N_{\rm s}\zeta_{\rm off}^2/4$, where $N_{\rm s}$ is the number of sources in the radio sample, here equal to 981 (see \citetalias{2017ApJ...845...89V}).  Hence the expected number of radio sources in our sample with the required alignment is $ N_{\rm f}N_{\rm s}\zeta_{\rm off}^2/4 = 2.53 \times 10^{-5}$. As pointed out in  \S  \ref{absence}, we should allow for the possibility that, due to possible distortion of the infrared galaxy disk isophotes, the separation between the infrared  blazar nucleus and the centroid of the spiral galaxy infrared isophotes could be as much as 26 milliarcseconds. So the correct probability to consider here is $1.0 \times 10^{-4}$.  While this is  a low value, since this is {\it a posteriori\/} statistics the argument against the background source hypothesis is not compelling.

\subsection{Conclusion}
Of the three arguments presented in this section, as in \citetalias{2017ApJ...845...89V}, we here again  reject the alignment argument of \S \ref{aposteriori} against the radio source being a background source  because it is {\it a posteriori\/}.
 The  remaining two arguments, based on the mid-infrared spectrum (\S \ref{infrared2}) and the good alignment of the jet axis in powerful jetted-AGN  in spirals with the spiral spin axis (\S \ref{misaligned}), combined with the high 1.67 GHz luminosity of PKS 1413+135, lead us to conclude  that it is not located in the spiral galaxy but is a background object.

\section{Other Points to be Considered}\label{other}
There are several other observations and points that need to be considered, which we cover in this section.
  
\subsection{The High Absorption of the Soft X-ray Flux in PKS\,1413+135}\label{xray2}

The HST  observations of \citet{2002AJ....124.2401P} show that the jetted-AGN PKS\,1413+135 has a very red color, $V-H= 6.9$\,mag. They also deduced from their HST and \textit{ASCA} observations that $N_{\rm H} = 4.6^{+2.1}_{-1.6} \times 10^{22}\;{\rm cm}^{-2}$. 
All of this is consistent with the high X-ray extinction, which implies $A_{\rm v} \sim 30$ mag, found by \citet{1992ApJ...400L..17S}. 

As pointed out by \citet{2002AJ....124.2401P}, the high X-ray extinction and much lower near-infrared extinction, which implies $A_{\rm v} \sim 14$, can easily be explained if the medium is patchy. Based on this assumption, they deduce a covering fraction $f=0.12^{+0.07}_{-0.5}$.   Furthermore \citep{2002AJ....124.2401P}, if this material is in the nuclear regions of the spiral galaxy, and if the jetted-AGN were in the spiral, one would expect to see a bright narrow 
 Fe K$\alpha$ line, but no such line is seen in their \textit{ASCA} spectrum, although the sensitivity is only sufficient to place an upper limit of 500\,eV on the equivalent width. The non-detection of this line is consistent with the jetted-AGN being a background source.

\subsection{Polarization in PKS\,1413+135}\label{lowradpol}
  
PKS\,1413+135 has low polarization at radio frequencies \citep{1999ApJ...512..601A}.  This has long been something of  a puzzle given its blazar and BL~Lac nature, since blazars and BL~Lac objects are well-known to be highly polarized  \citep[e.g.,][]{1993ApJ...416..519C,2013EPJWC..6106001W}.  But PKS\,1413+135 is being viewed through an edge-on spiral galaxy, so that  Faraday depolarization could explain this. Support for this hypothesis comes from both the infrared observations of \citet{1981Natur.293..714B}, who measured high infrared polarization in $H$-band (16\% $\pm$ 3\% in PA $175\degr \pm 5\degr$), and those of \citet{1992ApJ...400L..17S} who measured high infrared polarization in $K$-band (10.5\% $\pm$ 1.4\% in PA $170\degr \pm 1.4\degr$).

  \subsection{The Need for a Multi-zone Model}\label{multizone}

Like most blazars, the SED of PKS\,1413+135, shows two peaks: a synchrotron peak in the optical--ultraviolet  range and an inverse-Compton peak in the GeV energy range.  Many studies of AGN are based on the SED and use a single-zone model in which  the emission causing both the synchrotron peak and the inverse-Compton peak is assumed to originate from the same particle population.

The first VLBI images of blazars, namely 3C\,273 and 3C\,345, showed that they are one-sided jets with a flat-spectrum, optically thick, core at one end of a steep-spectrum, optically thin, jet \citep{1978Natur.276..768R}. This has turned out to be the case in most blazars. The higher-frequency radio emission is dominated by emission regions closer to the center of activity than the lower-energy emission. Thus it is reasonable to expect that the optical synchrotron emission from blazar jets originates closer to the core than the radio emission regions. \citet{2017A&A...598L...1K} compared the positions of the radio cores and jets of jetted-AGN with the optical positions measured with Gaia, and they found differences from less than 1\,mas up to 10\,mas. They also found  that while in some blazars the centroid of the optical emission lies on the  side of the radio core opposite to the jet, and hence closer to the central engine, as expected from the VLBI results discussed above, for a significant fraction of blazars the centroid of the optical positions lie further out along the jet than the nuclear radio jet, which was totally unexpected. It is clear, therefore, that in many blazars a single-zone model cannot explain both the synchrotron radio jet and the synchrotron optical jet.

We have seen in \S \ref{xray2} that the continuum  X-ray emission from PKS\,1413+135 comes from a region much smaller than the near-infrared continuum emission region. The single-zone model cannot, therefore, explain the beamed infrared and X-ray emission in the relativistic jet of PKS\,1413+135, and we see from Fig.\ref{map} that it likewise cannot explain the relativistically beamed  radio emission. Thus a multi-zone model is needed in all frequency ranges to explain the blazar emission from PKS\,1413+135.

\section{Conclusions}\label{discussion}

In view of the combination of observed properties, it is not surprising that the radio source PKS\,1413+135 has been so hard to understand. The circumstantial evidence supporting the hypothesis that it is located in the spiral galaxy is  persuasive  -- (i) the projected orthogonality of the radio jet to the galactic disk, (ii) the lack of multiple images of the radio core on arcsecond scales, and (iii) the close alignment between the blazar and the centroid of the near-infrared isophotes of  the spiral galaxy -- and therefore hard to set aside. But we have done so in the light of what we think  is compelling evidence to the contrary, based on the blazar nature and jet orientation of PKS\,1413+135, its variability Doppler factor, its superluminal motion, and its high luminosity coupled with the absence of any signatures of reprocessed radiation in the infrared spectrum.  The peculiar features of PKS\,1413+135 as a proposed member of the CSO  class further complicated the case. We have attempted to address all of these issues  comprehensively, and we hope that they have now all been set to rest.

We conclude with the following major findings:
\begin{enumerate}
\itemsep0em 
\item The jet axis of the radio source PKS\,1413+135 is  closely aligned with the line of sight.
\item The radio source PKS\,1413+135 is almost certainly located behind the spiral galaxy at redshift $0.247\, < \, z \, < \, 0.5$.
\item The spiral galaxy at $z = 0.247$ is a Seyfert 2 AGN powered by a $\sim 10^8 M_\odot$ SMBH.
\item The intervening spiral galaxy provides a natural host for a milli-lens that could be the cause of SAV. 
\item A multizone model is needed to explain the infrared, X-ray, and $\gamma-$ray emission, in addition to the radio emission, from PKS\,1413+135.
\item PKS\,1413+135 should not be classified as a CSO because its radio core is relativistically beamed toward the observer.
\end{enumerate}

The cause of SAV events has not yet been established beyond doubt. We are continuing our intensive campaign of multi-frequency observations of this remarkable object in order to establish whether or not SAV is indeed a gravitational milli-lensing phenomenon.

%\clearpage
\acknowledgments

We thank the anonymous referee for helpful comments on the original manuscript. We thank Jonathan Sievers and  Charles Steidel for useful discussions.
The OVRO 40\,m program was supported by NASA grants NNG06GG1G, NNX08AW31G, NNX11A043G, and NNX13AQ89G from 2006 to 2016 and NSF grants AST-0808050, and AST-1109911 from 2008 to 2014. The Submillimeter Array is a joint project between the Smithsonian Astrophysical Observatory and the Academia Sinica Institute of Astronomy and Astrophysics and is funded by the Smithsonian Institution and the Academia Sinica. T. H. was supported by Academy of Finland projects 317383 and 320085. W.M. acknowledges support from ANID projects Basal AFB-170002 and PAI79160080. R.R. acknowledges support from ANID Basal AFB-170002, and ANID-FONDECYT grant 1181620.   This work is based in part on archival data obtained with the \textit{Spitzer} Space Telescope, which was operated by the Jet Propulsion Laboratory, California Institute of Technology under a contract with NASA. Support for this work was provided by an award issued by JPL/Caltech.  The Combined Atlas of Sources with \textit{Spitzer} IRS Spectra (CASSIS) is a product of the IRS instrument team, supported by NASA and JPL. CASSIS is supported by the ``Programme National de Physique Stellaire'' (PNPS) of CNRS/INSU co-funded by CEA and CNES and through the ``Programme National Physique et Chimie du Milieu Interstellaire'' (PCMI) of CNRS/INSU with INC/INP co-funded by CEA and CNES.

This paper is dedicated to the memory of Neil Gehrels, who facilitated support that enabled us to bring the 40 m Telescope out of mothballs and commission it for the 15\,GHz monitoring program in support of {\it Fermi}-LAT, without which this work could not have been done.

\facilities{Fermi (LAT), Keck:I (LRIS), Mets{\"a}hovi Radio Observatory, OVRO:40m, SMA, Spitzer, UMRAO, VLBA}

%\clearpage
%\newpage
\appendix

\restartappendixnumbering
 
\section{Gravitational Lensing of a Point Source by a Point Mass}\label{pointmass}

The SMBH in the spiral galaxy is effectively a point mass since the Schwarzchild radius is $\leq 10^{-4} \times$ the Einstein radii we are considering here.  We will consider the simplest case of gravitational lensing by a point mass with no other masses involved.  The SMBH is embedded in a galaxy and so it exists in the presence of a smoothly varying background mass distribution, which will add convergence and shear terms to the potential,< but we do not consider these complications here since we are simply exploring this lensing situation, a detailed study of which is beyond the scope of this paper. 
 
\begin{figure}
    \centering
    \includegraphics[width=0.99\linewidth]{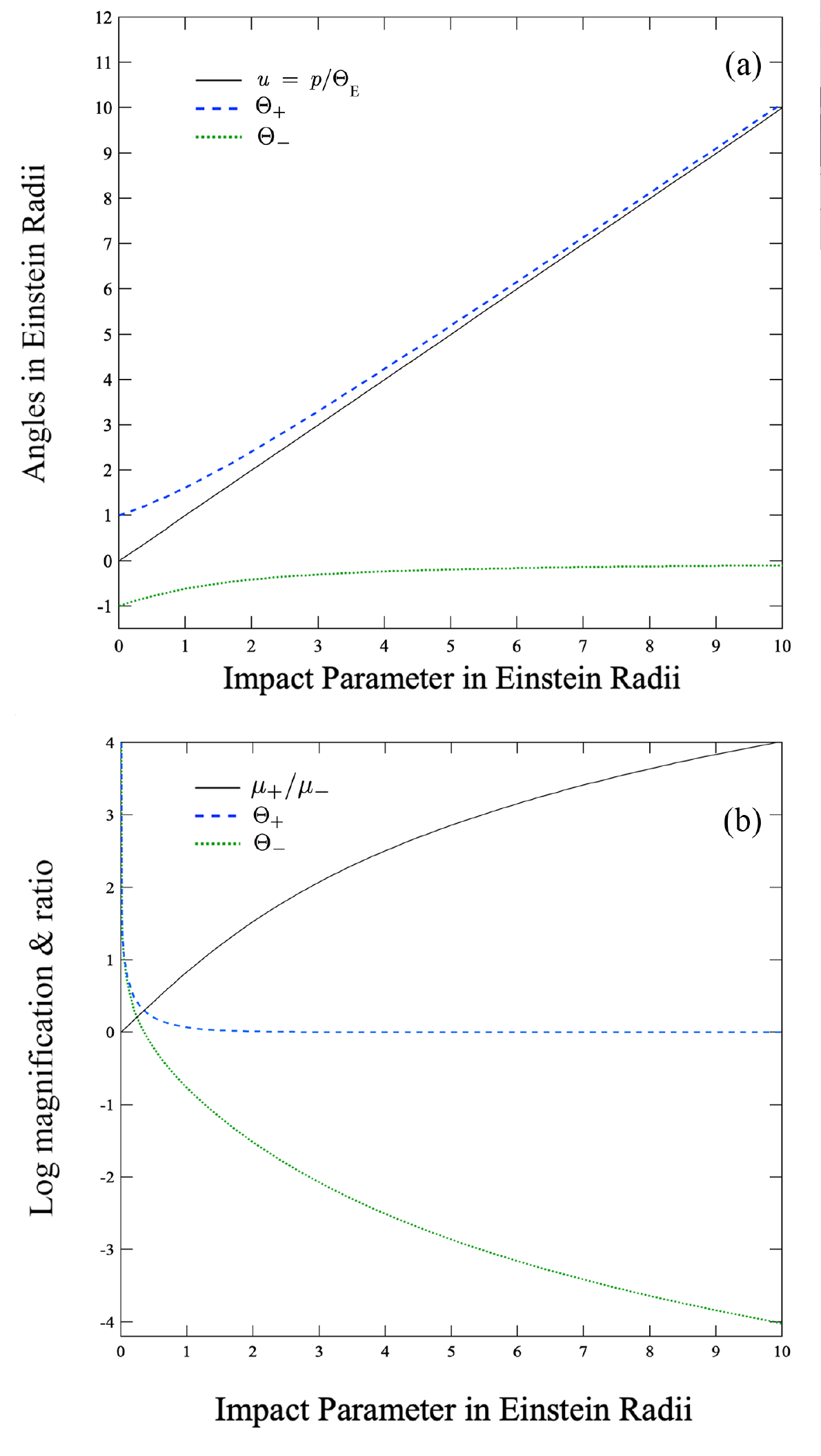}
    \caption{Separation and magnification  of the two images of a point source produced by a point mass gravitational lens. (a) Image position relative to the location of the point mass in units of the Einstein radius. The black line indicates the source position as it is moved from the location of the point mass out to 10 Einstein radii. The blue-dashed (upper) curve shows the position of the brighter image relative to the point mass, and the green-dotted (lower) curve shows the position of the fainter image relative to the point mass. (b)  Magnification of the two images and the magnification ratio as a function of the impact parameter in units of the Einstein radius.}
    \label{lenspars}
\end{figure}

Gravitational lensing of a point source by a point mass produces two images,  which we will designate as the ``Image+'' and ``Image$-$'' images. Image+    lies outside the Einstein radius,  and  Image$-$ lies inside the Einstein radius. We denote  the angle between the point mass and Image+, normalized by the Einstein radius, by $\Theta_+$;   and the angle between the point mass and Image$-$, normalized by the Einstein radius, by $\Theta_-$. We denote the magnification of Image+ by $\mu_+$ and the magnification of Image$-$ by $\mu_-$. As the normalized impact parameter $u \rightarrow \infty$,  $ \Theta_+ \rightarrow u, \; {\rm and}\; \Theta_- \rightarrow 0$, while $\mu_+ \rightarrow 1$, and $\mu_- \rightarrow 0$. The relative positions and magnifications of the two images  are given (see, e.g.,  \citealt{1964MNRAS.128..295R}; \citealt{1996astro.ph..6001N})
by
\begin{equation}\label{Theta}
    \Theta_{\pm} = {1 \over 2} \biggl(u \pm \sqrt{u^2+4}\biggr)
\end{equation}
and
\begin{equation}\label{mu}
    \mu_{\pm} = {u^2+2 \over {u\sqrt{u^2+4}}}\pm {1 \over 2} \, .
\end{equation}
In Fig.~\ref{lenspars} we illustrate the situation for impact parameters out to ten Einstein radii. Shown here are both the angular displacements of the images from the true source position, and the magnifications of the two images and the ratio of their flux densities.

Note that according to equation~(\ref{mu}), and hence by the ratio $\mu_+/\mu_-$, which is shown by the black curve in Fig.~\ref{lenspars}(b), the absence of a secondary image, which places a lower limit on this ratio,  enables us to place a lower limit on the impact parameter, $u$, measured in Einstein radii, and hence on the angle between the lines of sight to the lens and the source.

\restartappendixnumbering
\section{The Variability Doppler Factor}\label{appvar}

In a source emitting incoherent electron synchrotron radiation \citet{1957ApJ...125....1B} showed that if there is equipartition between the magnetic field and particle energy densities it is possible to determine the equipartition magnetic field, $B_{\rm eq}$, from the flux density, angular size, and distance of the source. They also showed that equipartition energy of the source is only slightly greater than the minimum energy possible for a source of incoherent synchrotron radiation producing the observed flux density.

Synchrotron self-absorption (SSA) in an incoherent electron synchrotron source provides a completely different method of determining the magnetic field, $B_{\rm SSA}$, which does not depend on the assumption of equipartition  \citep{1963Natur.199..682S,1963Natur.200...56W}.  $B_{\rm SSA}$ can be determined from the frequency, flux density, and angular size of a synchrotron source at the peak  frequency in a source showing SSA.

\begin{figure}
    \centering
    \includegraphics[width=0.99\linewidth]{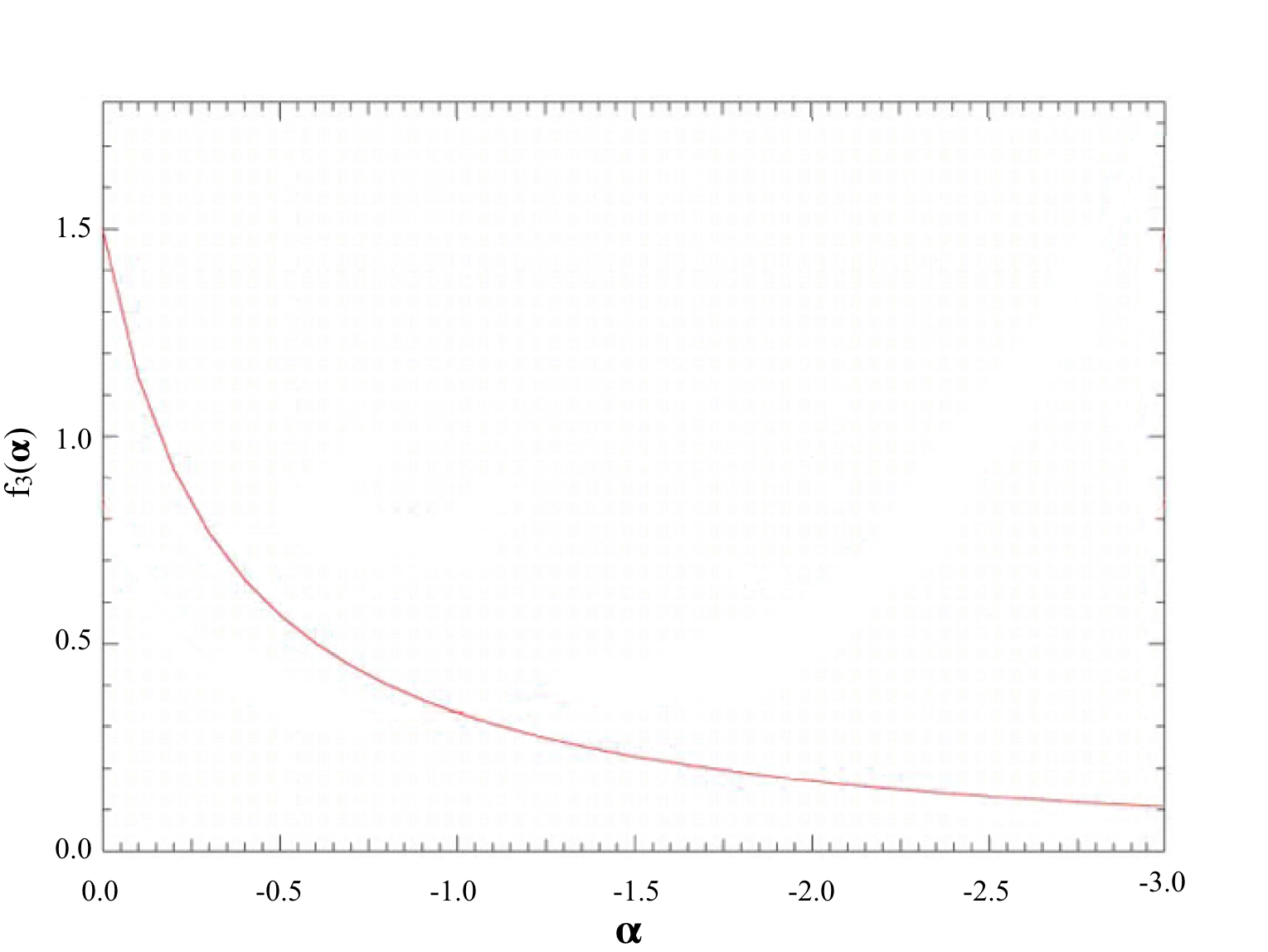}
    \caption{The spectral index function $f_3(\alpha)$}
    \label{f3alpha}
\end{figure}

From \citet{1968ARA&A...6..321S} we derive the following expression for the magnetic field $B$ in a source showing synchrotron self-absorption:
\begin{equation}\label{SSA}
B_{\rm {_{SSA}}} \le 1.0 \times 10^{-6} \ \biggl({S \over {\psi^2}}\biggr)^{-2} \, \nu^5 \, f_3(\alpha)^2 \, D \, ,
\end{equation}
where $B$ is in gauss, $S$ is the flux density in jansky at the peak of the spectrum, $\psi$ is the source angular diameter of a uniform brightness disk at frequency $\nu$ in arc seconds, $\nu$ is the frequency at the peak in the spectrum in megahertz, and  $f_3(\alpha)$ is shown in Fig.~\ref{f3alpha}. This spectral index function is given by equation~(15c) of \citet{1968ARA&A...6..321S}, noting that they use the convention $S \propto \nu^{-\alpha}$. $f_3(\alpha)$ describes the case for a ``tangled'' magnetic field, i.e., the magnetic field direction varies randomly throughout the emission region with  no preferred axis. The equality sign in equation~(\ref{SSA}) applies at the peak of the spectrum for a source showing SSA, the `$<$' applies for an optically thin source on the straight (unabsorbed) part of the spectrum at frequencies above the SSA turnover frequency.
 
We see from equation~(\ref{SSA}) that the magnetic field derived from synchrotron self-absorption depends on the inverse square of the flux density, the fourth power of the angular size, and the fifth power of the frequency of the spectral turnover.  Thus $B_{\rm {_{SSA}}}$  is poorly constrained by the observable properties of the source.

This led \citet{1977MNRAS.180..539S} to invert the problem by setting $B_{\rm SSA} = B_{\rm eq}$ and hence to define the equipartition angular size, $\psi_{\rm eq}$, which depends weakly on the source redshift and  flux density, and almost linearly on the peak frequency. Thus the equipartition angular size is a robust parameter that can be determined from observations with high accuracy in view of its relatively weak dependence on observable quantities. This makes it an extremely useful parameter in the study of the relativistic jets of blazars.

\citet{1977MNRAS.180..539S} determined the equipartition angular size for the case of an Einstein de-Sitter cosmology. For the $\Lambda$CDM cosmology we must use
the comoving coordinate distance:
\begin{equation}\label{rcos}
    r = {c \over H_0}\; \int\limits_0^z\, {{dz} \over {\sqrt{\Omega_{\Lambda}+\Omega_m(1+z)^3}}} \, .
\end{equation}

In the $\Lambda$CDM cosmology the equipartition angular size becomes:
\begin{equation}\label{psi1}
\psi_{_{\rm eq}} = 1.67 \times  r^{-{1 \over17}} S^{8 \over 17}  \nu^{-{{35+ 2\alpha} \over 34}}   (1+z)^{{15-2\alpha} \over 34}
 F(\alpha) \, ,
\end{equation}
where $r$ is in gigaparsecs,  and $F(\alpha)$ is given in \citet{1977MNRAS.180..539S}, noting that they use the convention $S \propto \nu^{-\alpha}$, and the other parameters are as in equation~(\ref{SSA}).

Similarly the equipartition brightness temperature, $T_{\rm eq}$  \citep{1994ApJ...426...51R}, is a robust observable in the sense that it depends relatively weakly on observable parameters and can be accurately determined from observations.  \citet{1994ApJ...426...51R} derived the equipartition brightness temperature for the case of an Einstein de-Sitter cosmology.  In the $\Lambda$CDM cosmological model the equipartition brightness temperature is given by:
\begin{eqnarray}\label{Teq}
T_{\rm eq} &=&5.75\times10^{11} \biggl[{{r\over {(1+z)}}}\biggr]^{2/17} F(\alpha)^{-2} \nonumber\\
 &\times&(1+z)^{(2\alpha - 
13)/17}S^{1/17}\bigl(10^3\nu\bigr)^{(1+2\alpha)/ 
17}{\rm 
K} \, .
\end{eqnarray}
Typical equipartition brightness temperatures in compact radio emission regions in AGN are $\sim 10^{11}$ K \citep{1994ApJ...426...51R,2018ApJ...866..137L}.
 
 If the observer or the emission region has peculiar velocity $v$ relative to the Hubble flow, and corresponding Doppler factor $D$, then the equipartition brightness temperature is
\begin{equation}\label{Teq1}
 T'_{\rm eq} = D T_{\rm eq} \, ,
\end{equation}
where the prime indicates a quantity measured in the observer's frame.

\citet{1977MNRAS.180..539S} showed for a small sample of radio sources, selected at the low frequency of 81.5 MHz so as to be radiating isotropically, and therefore having no peculiar motion relative to the Hubble flow, that the  compact emission regions are close to equipartition.
\citet{1994ApJ...426...51R} showed this for two large samples of sources \citep{1974MmRAS..78....1R,1975ApJ...202..596B}, also selected at low frequencies so as to be radiating isotropically, and introduced the equipartition Doppler factor:
\begin{equation}\label{Deq}
D_{\rm eq} = T_{\rm obs}/T_{\rm eq} \, ,
\end{equation}
where  $T_{\rm obs}$ is the observed brightness temperature as measured directly, e.g., by VLBI.  

An alternative approach to determining the Doppler factor of equation~(\ref{Deq}) through direct measurement of the angular size by interferometry is provided by the variability observed in blazars. 

The discovery that the quasars 3C\,273, 3C\,279 and 3C\,345  showed significant variability at radio frequencies on timescales less than a year \citep{1965Sci...148.1458D,1965Sci...150...63M}, led \citet{1966Natur.211..468R,1967MNRAS.135..345R} to suggest that in some quasars the emission region is moving at relativistic speed towards the observer. 

Consider motion of an emission region at a relativistic speed $v$ towards the observer at angle $\theta$ to the line of sight \citep[see, e.g.,][]{1979Natur.277..182S,1984RvMP...56..255B}. The Lorentz $\gamma$ factor is
\begin{equation}\label{gamma}
 \gamma= {1 \over {\sqrt{1-\beta^2}}} \, ,
\end{equation}
where  $\beta = v/c$. The Doppler factor is 
\begin{equation}\label{Dop}
D = {1 \over {\gamma (1-\beta \cos\theta)}} \, .
\end{equation}
Radiation emitted at frequency $\nu_{\rm em}$ will be observed at frequency 
\begin{equation}\label{nuobs}
\nu_{\rm obs} = D \nu_{\rm em} \, .
\end{equation}
The apparent transverse speed of the emission region is
\begin{equation}\label{vapp}
v_{\rm app} = { v \sin \theta \over {1 - \beta \cos \theta}} 
\end{equation}
and the observed flux density, $S_{\rm obs}$, is related to the emitted flux density, $S_{\rm em}$ at the emitted frequency $\nu_{\rm em}$ by
\begin{equation}\label{sobs}
S_{\rm obs} = S_{\rm em} D^{3-\alpha} \, .
\end{equation}

The maximum apparent transverse speed is $\gamma v$, which occurs at $\sin\theta = 1/\gamma$, where $D = \gamma$.  In addition, it should be noted that for objects moving at relativistic speeds ($v  \sim c$) equation~(\ref{vapp}) shows that $v_{\rm app} \rightarrow c$ as $\theta \rightarrow 90\degr$, and $v_{\rm app} \rightarrow 0$ as $\theta \rightarrow 0\degr$.  Thus any apparent superluminal transverse speed $c < v_{\rm app}< \gamma v$ will be seen at two different angles $\theta$.

The observed timescale of variability in blazars, $\tau$, can be used to determine physical conditions in the emission regions on the assumption that 
the proper diameter of the emission region, $\eta$, is less than the velocity of light times the proper time  of the variation, $\delta t$.  For a source at redshift $z$ moving at peculiar velocity $v$ relative to the Hubble flow we have
\begin{equation}\label{time}
\eta \le c\,\delta t=c\,{D\tau \over {1+z}} \, .
\end{equation}
If $\psi$ is the observed source angular diameter then we have
\begin{equation}\label{psi2}
\psi = {\eta(1+z) \over r} \le {c\, D \tau \over r} \, ,
\end{equation}
where $r$ is the comoving coordinate distance given by equation~(\ref{rcos}).  If we assume the proper diameter of the source is equal to the velocity of light times the proper time of the variation in equation~(\ref{time}), and apply the Rayleigh-Jeans law, the corresponding observed brightness temperature is
\begin{equation}\label{temp}
T_{\rm b} = 1.22\times 10^{11}\,{\Delta S \lambda^2 r^2 \over {D^2 \tau^2}}\;{\rm K} \, ,
\end{equation}
where  $\Delta S$ is the change in flux density in janskys,  $\tau$ is the observed time of the flux density variation in years, $\lambda$ is the observing wavelength in centimeters, and $r$ is the comoving coordinate distance in gigaparsecs given by equation~(\ref{rcos}).  \citet{1994ApJ...426...51R} showed that if $T_{\rm b}>T_{\rm eq}$ the total energy of the emission region scales as $T_{\rm b}^3$ and the ratio of the magnetic field energy density to the particle energy density scales as $T_{\rm b}^{8.5}$. Given that typical equipartition temperatures in AGN are $T_{\rm eq} \sim 10^{11}$ K and that many blazars show strong variability  at cm wavelengths on timescales of weeks,  we see from equation~(\ref{temp}) that significant Doppler boosting is required if we are to avoid very large inferred total energies and energy densities in blazars.  

We define the observed timescale of variability in blazars by 
\begin{equation}\label{tau}
\tau_{\rm var} = dt/d(\ln S)
\end{equation}
and assume that the timescale of the variation is equal to the light travel time across the component.
In the case where we use the variability to estimate the angular size to determine $T_{\rm obs} \; (= T_{\rm var})$ in equation~(\ref{Deq}), as seen in equation~(\ref{temp}) the Doppler factor enters twice more in the squared timescale that yields the solid angle used to determine the observed brightness temperature  and hence  the variability Doppler factor is
\begin{equation}\label{Dvar}
D_{\rm var} =  (T_{\rm var}/T_{\rm eq})^{1/3} \, .
\end{equation}

The determination of the angular size, and hence the brightness temperature through the variability of the source used in equation~(\ref{Dvar}) is not as reliable as the measurement of the size by interferometry used in equation~(\ref{Deq}), so this introduces additional uncertainty into the determination of the Doppler factor. However this is mitigated by the fact that $D_{\rm var}$ depends only on the cube root of $T_{\rm var}$.  This, and the fact that variability timescales are much easier to measure than angular sizes, makes the variability Doppler factor a reliable, powerful, and widely-used approach to measuring Doppler factors in blazars \citep{1999ApJS..120...95V,1999ApJ...521..493L, 2009A&A...494..527H,2015MNRAS.454.1767L,2017ApJ...846...98J,2018ApJ...866..137L}.  This is the approach that we adopt in \S \ref{varsav}.

\newpage

\bibliography{sample63}{}
\bibliographystyle{aasjournal}

\end{document}